\documentclass[12pt]{article}
\usepackage{amsmath,amssymb} 
\usepackage{subcaption}   

\usepackage{amsfonts}
\usepackage{enumerate}
\usepackage{ytableau}
\usepackage{hyperref}

\numberwithin{equation}{section}
\usepackage{cite}
\usepackage{bm}
\usepackage{tikz}
\usepackage{blkarray}  
\begin{document}

\def\im{\text{i}}
\def\eqa{\begin{eqnarray}}
\def\eqae{\end{eqnarray}}
\def\be{\begin{equation}}
\def\ee{\end{equation}}
\def\bea{\begin{eqnarray}}
\def\eea{\end{eqnarray}}
\def\ba{\begin{array}}
\def\ea{\end{array}}
\def\bd{\begin{displaymath}}
\def\ed{\end{displaymath}}
\def\eg{{\it e.g.~}}
\def\ie{{\it i.e.~}}
\def\Tr{{\rm Tr}}
\def\tr{{\rm tr}}
\def\>{\rangle}
\def\<{\langle}
\def\a{\alpha}
\def\b{\beta}
\def\c{\chi}
\def\del{\delta}
\def\e{\epsilon}
\def\f{\phi}
\def\vf{\varphi}
\def\tvf{\tilde{\varphi}}
\def\g{\gamma}
\def\h{\eta}
\def\j{\psi}
\def\k{\kappa}
\def\l{\lambda}
\def\m{\mu}
\def\n{\nu}
\def\w{\omega}
\def\p{\pi}
\def\q{\theta}
\def\r{\rho}
\def\s{\sigma}
\def\t{\tau}
\def\u{\upsilon}
\def\x{\xi}
\def\z{\zeta}
\def\D{\Delta}
\def\F{\Phi}
\def\G{\Gamma}
\def\J{\Psi}
\def\L{\Lambda}
\def\W{\Omega}
\def\P{\Pi}
\def\Q{\Theta}
\def\S{\Sigma}
\def\U{\Upsilon}
\def\X{\Xi}
\def\nab{\nabla}
\def\pa{\partial}
\newcommand{\lra}{\leftrightarrow}
\newcommand{\intbar}{\text{ (p.v.)}\int}

\newcommand{\bc}{{\mathbb{C}}}
\newcommand{\br}{{\mathbb{R}}}
\newcommand{\bz}{{\mathbb{Z}}}
\newcommand{\bp}{{\mathbb{P}}}
\def\ml{\mathcal{L}}
\newcommand{\cl}{{\cal L}}
\def\({\left(}
\def\){\right)}
\def\nn{\nonumber \\}

\newcommand{\red}{\textcolor[RGB]{255,0,0}}
\newcommand{\blue}{\textcolor[RGB]{0,0,255}}
\newcommand{\green}{\textcolor[RGB]{0,255,0}}
\newcommand{\cyan}{\textcolor[RGB]{0,255,255}}
\newcommand{\magenta}{\textcolor[RGB]{255,0,255}}
\newcommand{\yellow}{\textcolor[RGB]{255,255,0}}
\newcommand{\sky}{\textcolor[RGB]{135, 206, 235}}
\newcommand{\orange}{\textcolor[RGB]{255, 127, 0}}
\def\d{\operatorname{d}}
\def\ttbar{T$\overline{\text{T}}$ }
\def\Renyi{R$\acute{\text{e}}$nyi }
\def\Poincare{Poincar$\acute{\text{e}}$ }
\def\Banados{Ba$\tilde{\text{n}}$ados }
\def\Mobius{M$\ddot{\text{o}}$bius }
\def\fd{\mathfrak{d}}
\title{\textbf{The Holography of the 2D inhomogeneously deformed CFT}}
\vspace{14mm}
\author{Zhehan Li$^{1}$\footnote{202412601001@email.sxu.edu.cn}, Zhifeng Li$^{2}$\footnote{202101101227@email.sxu.edu.cn}, Jia Tian$^{1,3}$
\footnote{wukongjiaozi@ucas.ac.cn}
}
\date{}
\maketitle

\begin{center}
	{\it $^1$State Key Laboratory of Quantum Optics and Quantum Optics Devices, Institute of Theoretical Physics, Shanxi University, Taiyuan 030006, P.~R.~China\\
 \vspace{2mm}
        $^2$School of Physics and Electronics Engineering, Shanxi University, Taiyuan 030006, P.~R.~China\\
 \vspace{2mm}
		$^3$Kavli Institute for Theoretical Sciences (KITS),\\
		University of Chinese Academy of Science, 100190 Beijing, P.~R.~China 
	}
\vspace{10mm}
\end{center}

\makeatletter
\def\blfootnote{\xdef\@thefnmark{}\@footnotetext}  
\makeatother

\begin{abstract}
We systematically study inhomogeneous Hamiltonians in two-dimensional conformal field theories within the framework of the AdS/CFT correspondence by relating them to two-dimensional curved backgrounds. We propose a classification of inhomogeneous Hamiltonians based on the Virasoro coadjoint orbit. The corresponding bulk dual geometries are described by the generalized Ba$\tilde{\text{n}}$ados solutions, for which we introduce a generalized Roberts mapping to facilitate their study. Our classification provides previously underexplored classes of deformations, offering fresh insights into their holographic properties. Revisiting the well-known example of the M$\ddot{\text{o}}$bius Hamiltonian, we establish a connection to the 3D C-metric, which describes three-dimensional accelerating solutions. Furthermore, we extend our analysis to KdV-type asymptotic boundary conditions, revealing a broader class of solvable inhomogeneous Hamiltonians that are not linear combinations of Virasoro charges but instead involve KdV charges.
\end{abstract}

\baselineskip 18pt
\newpage

\tableofcontents

\section{Introduction and Summary}
In recent years, the study of inhomogeneous deformations has attracted considerable attention in both high energy and condensed matter physics. These deformations were initially introduced in spin chains or lattice models \cite{Vekic:1993, Vekic:1996,Gendiar:2008udd, Ueda:2008wx, Hikihara:2011mtb, Gendiar:2011} as smooth boundary conditions \cite{Vekic:1993, Vekic:1996} to suppress boundary or finite-size effects. A notable example of such effects is the inhomogeneity of the ground state in a homogeneous spin chain with boundaries. Remarkably, it was discovered that in a one-dimensional critical spin system, this inhomogeneity can be entirely removed by the Sine-Square Deformation (SSD) \cite{Hikihara:2011mtb}\footnote{Similar equivalences have been observed for other sinusoidal deformations \cite{Gendiar:2011}.}, which ensures that the ground state of the deformed open system becomes equivalent to that of a uniform periodic system \cite{Gendiar:2008udd, Hikihara:2011mtb}. This equivalence has been rigorously proven in the context of 1D free fermions \cite{Katsura:2011zyx}.

In \cite{Katsura:2011ss}, followed by \cite{Tada:2014kza,Ishibashi:2015jba,Ishibashi:2016bey}, the underlying mechanism of this equivalence is elucidated from the perspective of conformal field theories (CFTs). It was shown that the SSD Hamiltonian can be expressed as a special linear combination of Virasoro generators \eqref{hssd}, leading to a novel quantization scheme, termed dipolar quantization, in contrast to the traditional radial quantization of 2D CFTs defined on the complex plane. Dipolar quantization results in a continuous Virasoro algebra, implying that the deformed CFT possesses an infinite spatial extent and a continuous energy spectrum. Building on this, a one-parameter deformation, known as the \Mobius deformation, was introduced in \cite{Okunishi:2016zat, Wen:2016inm}, interpolating between the undeformed Hamiltonian $H_0$ and the SSD Hamiltonian $H_{\text{SSD}}$. 

By making use of the conformal transformation of 2D CFTs, a plethora of inhomogeneous deformations are introduced in \cite{Wen:2016inm}. These models are particularly solvable because the deformations can be undone by applying appropriate conformal transformations. Despite their solvability, people may wonder their physical significance and potential applications. It was suggested in \cite{Wen:2018vux} that these deformations provide a framework for studying quantum quenches and CFTs defined on curved spacetimes. Subsequently, they were also employed to investigate Floquet or driven CFTs \cite{Wen:2018agb}, offering a platform to explore statistical or quantum physics beyond equilibrium. For instance, a fundamental question concerns how a system heats up, which is intimately related to the concepts of integrability, ergodicity, and quantum chaos. Heating dynamics can be modeled using a Floquet operator $F=e^{-\tau_0 H_0}e^{-\tau_1 H_{\text{SSD}}}$ \cite{Wen:2018agb,Fan:2019upv,Lapierre:2019rwj} or a quantum quench  $\rho(t)=e^{\im t H_{\text{\Mobius}}} \rho(0) e^{-\im t H_{\text{\Mobius}}}$ \cite{Goto:2021sqx}. By computing correlation functions or entanglement entropy, heating and non-heating phase diagrams have been obtained. From the Heisenberg perspective, the Floquet operator or the quantum quench implements a specific \Mobius conformal ($SL(2,\mathbb{R})$) transformation. Later works \cite{Han:2020kwp,Wen:2020wee} further explored the phase diagrams of general general $SL(2,\mathbb{R})$ deformation. In \cite{Fan:2020orx,Lapierre:2020ftq} more general inhomogenous Hamiltonian constructed as arbitrary linear combinations of Virasoro charges, is studied. For studies on other perspectives on the inhomogenous Hamiltonians, see \cite{Maruyama:2011njv,Lapierre:2024lga,Goto:2023yxb,Caputa:2020mgb,Moosavi:2019fas, MacCormack:2018rwq,Tamura:2017vbx,Tada:2017wul,Kishimoto:2018ekq,Hotta:2018hkq,Tada:2019rls,Liu:2020rxi,Lapierre:2020roc,Andersen:2020xvu,Das:2021gts,Hotta:2021ydu,Wen:2021mlv,Eguchi:2021duy,deBoer:2021zlm,Kawano:2022prj,Choo:2022lgm,Wen:2022pyj}.

While there are infinite ways to introduce deformations,  in this work, we aim to investigate inhomogeneous Hamiltonians in a more systematic manner using the framework of the AdS/CFT correspondence \cite{Maldacena:1997re,Witten:1998qj,Gubser:1998bc}. The key idea is to leverage the correspondence between inhomogeneous Hamiltonians and 2D curved spacetimes, allowing us to reformulate the problem in terms of the holography of CFTs on curved spacetimes. When the chiral and anti-chiral sectors are deformed in the same way, the resulting curved spacetime is static and fully characterized by a deformation function (also referred to as an envelope function). For each choice of the deformation function, we can construct a family of static asymptotically AdS geometries \eqref{family}. These AdS geometries preserve residual conformal symmetries, under which the deformation function transforms as an adjoint vector of the Virasoro group. Using the theory of the Virasoro group, the deformation functions can be classified into four distinct classes corresponding to Virasoro coadjoint orbits. The static geometries describe either the vacuum state or black hole states of the CFT. Beyond these, conformal transformations can be applied to obtain more general (often time-dependent) geometries \eqref{general}, which serve as holographic duals for other excited states. To study these general geometries \eqref{general}, we derive generalized Roberts mappings that transform them into the \Poincare AdS$_3$ geometry. Compared to other holographic studies \cite{Goto:2023wai,Das:2022pez,Erdmenger:2021wzc,Das:2024lra,deBoer:2023lrd,Kudler-Flam:2023ahk,Nozaki:2023fkx,Bernamonti:2024fgx,Mao:2024cnm,Jiang:2024hgt,Miyata:2024gvr,Bai:2024azk}, our approach offers several advantages and new features: 1) In some existing works (e.g., \cite{Goto:2023wai,Das:2024lra}), the proposed geometries are \Banados solutions, which are dual to CFTs on flat spacetimes. However, inhomogeneous Hamiltonians should correspond to CFTs on curved spacetimes. Consequently, \Banados geometries do not directly describe inhomogeneous Hamiltonians; instead, a non-uniform cutoff and a redefinition of the radial coordinate are required. This redefinition introduces ambiguities in the interpretation of the radial coordinate; 2) After redefining the radial coordinate, the bulk metric is no longer in Fefferman-Graham (FG) gauge, making it challenging to derive the boundary energy-momentum tensor from the metric. In contrast, the bulk geometries \eqref{family} and \eqref{general} are naturally in FG gauge; 3) The asymptotic boundary conditions and symmetries can be systematically analyzed using the Chern-Simons formulation of AdS$_3$ gravity; 4) While the general geometries \eqref{general} have been considered in previous studies (e.g., \cite{Jiang:2024hgt,deBoer:2023lrd}), the generalized Roberts mapping was not provided.

\subsection*{Summary of our results}
In this work, we proposed a systematic holographic framework for studying 2D inhomogeneously deformed CFTs. We demonstrated how deformation functions can be classified into four distinct classes \eqref{class} using the Virasoro coadjoint orbit. Among these, we introduced representatives \eqref{example} for the two previously overlooked classes: the $T_{\Delta,n}$ and $\tilde{T}_{n,\pm}$ deformations. These deformations provide new insights into the holographic duals of inhomogeneously deformed CFTs. We showed that these deformed CFTs are dual to bulk geometries described by \eqref{family} and \eqref{general}, and we provided the explicit coordinate transformations \eqref{mapping} that relate these geometries to the \Poincare AdS$_3$ spacetime. This transformation generalizes the Roberts mapping to states with arbitrary energy-momentum tensor expectation values on an arbitrary curved boundary spacetime. Using the CS formulation, we carefully analyzed the asymptotic boundary conditions and their corresponding asymptotic symmetries. We pointed out that prior holographic studies implicitly assumed Dirichlet boundary conditions. By adopting this choice, we derived the asymptotic symmetries for general deformation functions, thereby generalizing previous results for the \Mobius and SSD deformations. Furthermore, by imposing KdV-type boundary conditions, we uncovered a new class of solvable inhomogeneous Hamiltonians that are not simple linear combinations of Virasoro charges but instead involve KdV charges.  Within this framework, we computed key physical quantities, including the holographic mass, black hole temperature and entropy, the Euclidean on-shell action, and the holographic entanglement entropy for the general bulk geometry \eqref{general}.

We analyzed explicit examples within each class of deformation:
\begin{enumerate}
    \item \textbf{\Mobius and SSD Deformations:} Revisiting the \Mobius and SSD deformations, we reproduced known results from the literature. Additionally, we observed that the dual geometry of a specific q-\Mobius deformation is identical to the Type-I 3D C-metric. This connection provides a more physical understanding of both the q-\Mobius deformation and the 3D C-metric.
     \item \textbf{$T_{n,\Delta}$ Deformation:} For the $T_{n,\Delta}$ deformation, where the deformation function has simple zeros, we found that the dual geometry can be interpreted as multiple copies of the \Poincare geometry glued together at the zeros of the deformation function. In this case, the black hole solutions are partitioned into subregions by disconnected black hole horizon segments. Interestingly, within each subregion, the horizon touches the AdS boundary, leading to divergences in the entropy contribution. However, we proposed a resolution: by considering the orientation and assigning negative entropy to subregions with opposite orientation, the total black hole entropy becomes finite and diffeomorphism-invariant. This interpretation suggests that the solution represents a non-equilibrium black hole ``emitting'' smaller black holes that carry fractions of the total energy and entropy, or alternatively, a black-hole-anti-black-hole pair.
    \item \textbf{$\tilde{T}_{n,\pm}$ Deformation:} For the $\tilde{T}_{n,\pm}$ deformation, which can be understood as a perturbation of the SSD deformation, we observed that both the effective size and black hole entropy diverge. Since there is no finite orbit invariant in this class, we concluded that there is no canonical way to regularize the geometry. However,  we found that the effective length and black hole entropy, though divergent, depend on an orbit invariant, which could have a physical interpretation.
    
    \item \textbf{KdV Hamiltonians:} Finally, we studied the level-one KdV Hamiltonian. In the $p \to 0$ limit, the Hamiltonian reduces to the \Mobius Hamiltonian. However, we emphasized that under KdV boundary conditions, the holographic mass differs from the Hamiltonian. We derived a fully analytic expression for the holographic entanglement entropy of a one-interval subsystem, providing a new solvable model for studying entanglement in inhomogeneous systems.
\end{enumerate}

\section{Preliminary}
\label{sec:pre}
In this section, we begin by reviewing inhomogeneously deformed CFTs and their connections to CFTs defined on curved spacetimes. Subsequently, we elucidate how Virasoro coadjoint orbits can be utilized to classify inhomogeneous Hamiltonians. Finally, we present an overview of the holographic descriptions of the inhomogeneous CFTs.
\subsection{Inhomogeneously deformed CFTs and 2D CFTs on curved spacetimes.}
In this work, we consider a 2D CFT defined on a circle of length $2\pi$ with a periodic boundary condition \footnote{For the choice of open boundary conditions, non-trivial boundary effects may arise \cite{Liu:2023tiq}. }. For a flat 2D spacetime with the metric $ds^2=-dt^2+d\varphi^2$, the Hamiltonian of the CFT is given by
\be 
H=\int_0^{2\pi}\frac{d\varphi}{2\pi}h(\varphi),\quad h(\varphi)\equiv(T(\varphi+t)+\bar{T}(\varphi-t))|_{t=0}, \label{undeH}
\ee 
where $h(\varphi)$ is the energy density, and $T(\varphi+t)$ and $\bar{T}(\varphi-t)$ represent the renormalized chiral and anti-chiral energy-momentum tensors, respectively. In terms of the generators $\{L_n\}$ of the Virasoro algebra, the Hamiltonian \eqref{undeH} can also be expressed as
\be 
H=L_0+\bar{L}_0-\frac{c}{12},
\ee 
where $c$ is the central charge of the CFT and $E_0=-\frac{c}{12}$ is the vacuum (Casimir) energy arising from the finite size nature of the system.

An inhomogeneous Hamiltonian is defined through an envelope function (or deformation function) as
\be 
H_f=\int_0^{2\pi}\frac{d\varphi}{2\pi}f(\varphi)\(T(\varphi)+\bar{T}(\varphi)\), \label{inhomo}
\ee 
where we take $f(\varphi)$ to be periodic in accordance with the choice of the periodic boundary condition. Expanding the envelope function into the Fourier modes:
\be 
f(\varphi)=\sum_{n}f_n e^{\im n \varphi},
\ee 
we find that the inhomogeneous Hamiltonian is a general linear combination of the Virasoro algebra charges:
\be 
H_f=\sum_{n\in\mathbb{Z}}f_n(L_n+\bar{L}_n)-\frac{f_0 c}{12}.
\ee 
Thus, $H_f$ generates a specific conformal transformation.
Since $H_f$ and $H$ do not commute generically, a primary state of the original CFT is not necessarily a primary state of the deformed CFT. For instance, the vacuum state of the original CFT remains the vacuum state of the deformed state only when the deformation is restricted to the global sector of the Virasoro algebra, namely the $SL(2, R)$ algebra spanned by $L_0, L_{\pm 1},\bar{L}_0,\bar{L}_{\pm 1}$. The most well-studied example in this context is the \Mobius deformation \cite{Tada:2014kza,Ishibashi:2015jba,Ishibashi:2016bey,Okunishi:2016zat,Wen:2016inm}, which corresponds to the following envelope function and  inhomogeneous Hamiltonian:
\bea  
f_{\text{\Mobius}}(\varphi)&=&1-\tanh(2\theta)\cos(\varphi),\quad \theta\in\mathbb{R},\\
H_{f_{\text{\Mobius}}}&=&L_0+\bar{L}_0-\frac{\tanh(2\theta)}{2}(L_{-1}+L_1+\bar{L}_{-1}+\bar{L}_1)-\frac{c}{12}.
\eea 
Taking the limit $\theta\rightarrow\infty$, the \Mobius deformation reduces to the SSD deformation \cite{Gendiar:2008udd,Hikihara:2011mtb}:
\bea 
&&f_{\text{SSD}}=2\sin^2\frac{\varphi}{2},\\
&&H_{f_{\text{SSD}}}=L_0+\bar{L}_0-\frac{1}{2}(L_{-1}+L_1+\bar{L}_{-1}+\bar{L}_1)-\frac{c}{12}. \label{hssd}
\eea 
The \Mobius	deformation can be generalized to the q-\Mobius deformation with
\bea 
f_{\text{q-\Mobius}}(\varphi)&=&1-\tanh(2\theta)\cos(q\varphi),\quad \theta\in\mathbb{R},\quad q\in \mathbb{N}^+,\\
H_{f_{\text{\Mobius}}}&=&L_0+\bar{L}_0-\frac{\tanh(2\theta)}{2}(L_{-q}+L_q+\bar{L}_{-q}+\bar{L}_q)-\frac{c}{12}.
\eea 
The inhomogeneous Hamiltonians of the form \eqref{inhomo} have been studied for various choices of the envelope function $f(\varphi)$. Our goal is to systematically study these deformations with the help of holography. The main observation is that the envelope function can be classified by the Virasoro coadjoint orbits. Roughly speaking, $f(\varphi)$ can be categorized into three types based on its zeros: (1) $f(\varphi)$ has no zeros and is sign-definite; (2) $f(\varphi)$ has double zeros and is semi-sign-definite; and (3) $f(\varphi)$ has simple zeros and alternates between positive and negative values. Notably, the \Mobius deformation belongs to the first type, while the SSD deformation belongs to the second type. More precisely, there are four classes of Virasoro coadjoint orbits:
\be 
S^1,\quad PSL^{(n)}(2,\mathbb{R}),\quad T_{n,\Delta},\quad \tilde{T}_{n,\pm},\quad (n\in \mathbb{N},\Delta\in \mathbb{R}-{0} ). \label{class}
\ee 
in the terminology of Witten \cite{Witten:1987ty}. Most previous studies, including those on the \Mobius and SSD or general $SL(2,R)$ deformations, have focused on the first two classes. To study the inhomogeneous deformations in the other two classes, we introduce the representative envelope functions
\be  
f_{T_{n,\Delta\equiv 4\pi b}}(\varphi)=\frac{2}{nF}\cos(\frac{n\varphi}{2})\(\sin(\frac{n\varphi}{2})+\frac{b}{n}\cos(\frac{n\varphi}{2})\),\quad f_{\tilde{T}_{n,\pm}}(\varphi)=\frac{1}{H_{n,q}}\sin^2\frac{n\varphi}{2}, \label{example}
\ee 
with 
\bea  
F(\varphi)&=&\cos^2\frac{n\varphi}{2}+\(\sin(\frac{n\varphi}{2})+\frac{2b}{n}\cos(\frac{n\varphi}{2})\)^2,\\
H_{n,q}&=&1+\frac{q}{2\pi}\sin^2(\frac{n\varphi}{2}),\quad q=\pm 1.
\eea   
We refer to the corresponding deformations as the $T_{n,\Delta}$-deformation and $\tilde{T}_{n,\pm}$-deformation, respectively. It has been shown in \cite{Balog:1997zz} that these two choices are the most general representatives of the orbits $T_{\Delta,n}$ and $\tilde{T}_{n,\pm }$. A distinguishing feature of these orbits is that they do not admit a constant representative. In other words, unlike the \Mobius deformation, there is no diffeomorphism that transforms the envelope function $f$ into a constant. 

This observation also provides new insight into why \Mobius deformation and SSD deformation behave so differently: although both belong to the $PSL^{(n)}(2,\mathbb{R})$ class, the SSD deformation can be understood as the $\Delta=0$ limit of the $T_{n,\Delta}$-deformation, while \Mobius deformation shares more similarity with the $S^1$ deformation. Additionally, we propose a setup to study the holography of 2D CFTs on curved spacetimes. Compared to the previous works, our setup has the advantage of allowing a proper study of asymptotic symmetries, asymptotic boundary conditions, and Euclidean on-shell action.

 First, let us explain the relationship between inhomogeneously deformed CFTs and CFTs defined on curved spacetimes. This connection arises from the fact that Hamiltonian generates time translation. Consequently, the inhomogeneous Hamiltnoian $H_f$ can be interpreted as the Hamiltonian of a CFT defined on the 2D spacetime with a curved metric:
\be 
ds^2_f=-f^2(\varphi)dt^2+d\varphi^2. \label{2dcurved}
\ee 
Thus, there is a one-to-one correspondence between an inhomogeneous Hamiltonian of the form \eqref{inhomo} and a curved 2D metric. Moreover, for any such curved metric, there exists a family of AdS$_3$ bulk solutions, as demonstrated below. 
\subsection{Virasoro coadjoint orbits and their holographic descriptions}
 We consider the family of static solutions \cite{Skenderis:1999nb,Perez:2016vqo} given by
\bea 
ds^2&=&\frac{dz^2}{z^2}+\frac{1}{z^2}(-f^2 dt^2+d\varphi^2)+\frac{u d\varphi^2+f(uf-2f'')dt^2}{2}\nonumber \\
&\quad &+z^2\frac{u^2d\varphi^2-(uf-2f'')^2dt^2}{16}, \label{family}
\eea 
where $f$ and $u$ are periodic functions of $\varphi\in[0,2\pi]$. As we approach the boundary with $z\rightarrow 0$, we find that the boundary metric exactly reduces to the desired form \eqref{2dcurved}:
\be 
ds^2\sim -f^2 dt^2+d\varphi^2.
\ee 
The bulk metric exhibits residue diffoemorphims \cite{Perez:2016vqo} $(z,\varphi)\rightarrow (\tilde{z},\tilde{\varphi})$, where $\tilde{\varphi}(\varphi)$ can be identified as the Virasoro group. Thus, we can still refer to these residue diffeomorphisms as conformal transformations. Under this transformation, the metric retains the general form \eqref{family} but with replacements $f\rightarrow \tilde{f},\, u\rightarrow \tilde{u}$. In particular, $f(\varphi)\partial_\varphi$ transforms as an adjoint vector field \eqref{vector}, while $ud\varphi^2$ transforms as a coadjoint vector field \eqref{covector} of the Virasoro group. The Einstein equations are equivalent to the stabilizer condition:
\be 
\mathcal D f=0,\quad \mathcal{D}\equiv (\partial_\varphi u)+2u\partial_\varphi-2\partial^3_\varphi\, \label{stabEqn}
\ee 
This is called the stabilizer condition because, under the infinitesimal conformal transformation generated by the solution of \eqref{stabEqn}, the adjoint vector field $f(\varphi)$ and the coajoint vector field $u(\varphi)$ remain unchanged. Therefore, the space of the bulk solutions is described by the diffeomorphic equivalence class of $u(\varphi)$, known as the coadjoint orbit of the Virasoro group, with $f(\varphi)$ being the stabilizer or little group of the orbit. The classification problem of the Virasoro coadjoint orbits has been extensively studied in \cite{Witten:1987ty,Balog:1997zz,Lazutkin:1975,Kirillov:1981,Segal:1981,Ovsienko:1990,Gorsky:1991}, and we provide a brief review in Appendix \ref{appendix:virasoro}. 

The third-order differential equation \eqref{stabEqn} can be integrated to yield an integral constant:
\be 
uf^2-2ff''+(f')^2=\fd, \label{orbitconst}
\ee 
which is invariant under the conformal transformation.
 To verify this, we can apply the following rules of transformations of the adjoint vector and coadjoint vector:
\bea 
&&\varphi\rightarrow \tilde{\varphi},\label{diff}\\
&&\tilde{u}(\tilde{\varphi}(\varphi))=\frac{1}{\tilde{\varphi}'^2}(u(\varphi)+2\{\tilde{\varphi},\varphi\}), \label{covector} \\ 
&&\tilde{f}(\tilde{\varphi}(\varphi))=\tilde{\varphi}' f(\varphi). \label{vector}
\eea  
Additionally, there is a symmetry under time dilation:
\bea  
&&t\rightarrow \lambda t,\\
&&f \rightarrow \lambda^{-1}f,\\
&&\fd \rightarrow \lambda^{-2}\fd.
\eea 
Later, we will demonstrate that when the orbit constant is positive $(\fd>0)$, the bulk solution possesses a horizon, and the temperature associated with this horizon is given by $T=\sqrt{\fd}/2\pi$, consistent with the time dilation symmetry, as in the Euclidean signature, the time coordinate is interpreted as the inverse of the temperature.

In this work, we focus on the static solutions given in \eqref{family}, as we assume that the chiral and anti-chiral sectors of the 2D CFT are deformed by the same deformation function. For more general deformations, for instance, see \cite{Fan:2020orx}, the bulk solution typically becomes time-dependent and can be described by the general metric \eqref{general}. We note that these general bulk solutions are also applicable to the study of Floquet or driven CFTs, which we plan to investigate in the context of inhomogeneous deformations in future work.

\section{The holography of CFTs on curved spacetimes}
\label{sec:asy}
In this section, we first address potential subtleties in defining the holographic framework for CFTs on curved spacetimes. We then proceed to discuss the asymptotic boundary conditions in our setup. Specifically, we demonstrate how the deformed Virasoro symmetry derived in \cite{Ishibashi:2016bey,Okunishi:2016zat} can be obtained from our holographic description.

\subsection{Review of the \Banados solutions}
Let us begin by briefly reviewing the holographic description of a 2D CFT on flat spacetime, following \cite{Banados:1998gg}. We consider a family of asymptotic AdS$_3$ solutions known as the \Banados solutions, which represent the most general solutions with flat asymptotic boundaries. The three-dimensional metric takes the general form
\bea 
ds^2=\frac{dr^2}{r^2}+\frac{1}{4}u_+ dw_+^2+\frac{1}{4}u_-dw_-^2+(r^2+\frac{u_+u_-}{16r^2})dw_+dw_-, \label{general_metric}
\eea 
where $u_+(w_+)$ and $u_-(w_-)$ are arbitrary functions of the light-cone coordinates $w_\pm=\varphi\pm t$, respectively. There exists a family of three-dimensional coordinate transformations, known as Roberts (or \Banados) mappings \cite{Roberts:2012aq} $(r,w_\pm)\rightarrow (r',w_\pm')$, under which the metric retains its form but with new functions $u_\pm'$. These transformations correspond to asymptotic symmetries \cite{Brown:1986nw}, mapping one AdS solution to another. As we approach the AdS boundary by taking $r=\Lambda \rightarrow \infty$, the metric reduces to a 2D flat metric $ds^2\sim  dw_+dw_-$, and the coordinate transformations can be identified with the 2D conformal transformations $w'_+(w_+)$ and $w'_-(w_-)$ in the CFT defined on this flat background. Furthermore, a state in the 2D CFT is dual to a \Banados solution, with the functions $u_\pm$ are related to the expectation values of the energy-momentum tensors via
\be 
u_+=-\frac{24}{c}\langle T(w_+)\rangle,\quad u_-=-\frac{24}{c}\langle \bar{T}(w_-)\rangle.
\ee 
In general, \Banados solutions are expected to be smooth. However, singular \Banados solutions can also have CFT interpretations. For example, conical AdS geometries are dual to scalar primary states, and holographic local quench geometries \cite{Nozaki:2013wia} are dual to specific local quantum quenches. Another subtlety arises when the bulk geometry is smooth, but the asymptotic boundary metric exhibits singularities, as in the \Banados solutions considered in \cite{Hung:2011nu} for computing holographic \Renyi entropy. A rigorous and complete study of \Banados solutions with singularities in the bulk or at the asymptotic boundary remains a significant challenge. Similar subtleties regarding singularities persist in the bulk solutions \eqref{family}, which we will not address in this work. 

Despite these subtleties, we assume that a  state of the 2D CFT defined on the curved metric \eqref{2dcurved} is dual to a bulk solution of the form \eqref{general}. The conformal transformation $\varphi\rightarrow \tilde{\varphi}$, or the more general diffeomorphism \eqref{mapping}, maps one state to another, even though the conformal transformation may alter the curvature of the asymptotic 2D metric. The allowed coordinate transformations are determined by the choice of asymptotic boundary conditions, which we will elaborate on below.
 
\subsection{The asymptotic boundary conditions}
\label{sec:boundary}
The study of asymptotic boundary conditions plays a crucial role in the holographic description. In particular, specifying these conditions is a fundamental step in establishing the AdS/CFT correspondence, as they determine the space of solutions and the associated asymptotic symmetries. In the context of AdS$_3/$CFT$_2$ duality it is usually convenient to conduct this analysis in the Chern-Simons (CS) formalism \cite{Achucarro:1986uwr,Witten:1988hc}. At least classically, the Einstein's gravity theory in AdS space admits a representation as the difference of two $SL(2,\mathbb{R})$ CS actions\footnote{The convention is:
\bea 
&&T_1=\left(
\begin{array}{cc}
 0 & 0 \\
 1 & 0 \\
\end{array}
\right),\quad T_0=\left(
\begin{array}{cc}
 \frac{1}{2} & 0 \\
 0 & -\frac{1}{2} \\
\end{array}
\right),\quad T_{-1}=\left(
\begin{array}{cc}
 0 & -1 \\
 0 & 0 \\
\end{array}
\right), \\
&&\kappa_{ab}=\langle T_aT_b\rangle=\begin{pmatrix}
	0&0&-1\\
	0&\frac{1}{2}&0\\
	-1&0&0
\end{pmatrix}. \nonumber
\eea}:
\bea 
&&I=I_{\text{CS}}[A^+]-I_{\text{CS}}[A^-],\\
&&{I}_{\text{CS}}[A]=\frac{\k}{4\pi}\int_\mathcal{M}\langle A\wedge d A+\frac{2}{3}A\wedge A\wedge A\rangle, \label{CSaction}
\eea 
where $\mathcal{M}$ represents a 3D spacetime with a ``radial" coordinate $r$ and two boundary coordinates $(t,\varphi)$, typically parameterizing a cylinder or torus. It is often advantageous to decompose the gauge field into spatial and temporal components: $A^\pm=A_i^\pm dx^i+A_t^\pm dt$, then the CS action \label{CSaction} can be expressed as
\be 
I_{CS}[A^\pm]=-\frac{\kappa}{4\pi}\int dt d^2x \epsilon^{ij}\langle A_i^\pm \dot{A}_j^\pm-A_t^\pm F_{ij}^\pm\rangle+B_\infty^\pm,
\ee  
where $\epsilon^{ij}$ is the antisymmetric tensor, and $F_{ij}^\pm \equiv \partial_i A_j^\pm - \partial_j A_i^\pm + [A_i^\pm, A_j^\pm]$ denotes the spatial field strength. The boundary terms $B_\infty^\pm$ are added to ensure proper variation of the action by canceling potential surface terms.
Following foundational works on asymptotic boundary conditions in CS formalism \cite{Coussaert:1995zp,Henneaux:2013dra,Bunster:2014mua}\footnote{For complete boundary condition classification, consult \cite{Grumiller:2016pqb}.}, we adopt the radial gauge parametrization:
\be 
A^\pm=g_\pm^{-1}(d+\mathcal{L}^\pm)g_{\pm},\quad g=e^{\pm (\log r) T_0}
\ee 
with $\mathcal{L}=\mathcal{L}_td t+\mathcal{L}_\varphi d \varphi$ and
\bea 
&&\cl_\varphi^\pm=T_{\pm 1}-\frac{1}{4}u_\pm T_{\mp 1}; \quad \cl_t^\pm=\pm \Lambda^\pm [f^\pm],\\
&&\Lambda^\pm [f^\pm]=f^\pm(T_{\pm 1}-\frac{1}{4}u_\pm T_{\mp 1})\mp {f^\pm}'T_0+\frac{1}{2}{f^\pm}''T_{\mp1},
\eea 
where the superscript $'$ denotes the derivative with respect to $\varphi$.
When $f^\pm =1$, the gauge field configuration reduces to \Banados's original solution \cite{Banados:1998gg}. The  flatness condition of the gauge fields yields the dynamical equations:
\begin{equation}\label{eom}
\begin{aligned}
\dot{u}_\pm &= \pm \mathcal{D}^\pm f^\pm, \\
\mathcal{D}^\pm &:= \partial_\varphi u_\pm + 2u_\pm \partial_\varphi - 2\partial_\varphi^3.
\end{aligned}
\end{equation}
For time-independent configurations ($\partial_t f^\pm = \partial_t u_\pm = 0$), these field equations exactly reproduce the stabilizer condition \eqref{stabEqn}. The corresponding bulk spacetime metric, constructed through the bilinear form
\begin{equation}
ds^2 = \frac{1}{2}\langle (A^+ - A^-)_\mu (A^+ - A^-)_\nu \rangle dx^\mu dx^\nu,
\end{equation} 
precisely matches the family of solutions \eqref{family}. Time-dependent generalizations of these solutions are detailed in Appendix \ref{appendix:general}.
 
The CS action variation yields boundary contributions through integration by parts:
\begin{equation}\label{bdyvar}
\delta I_{\text{CS}}^\pm = \pm \frac{\kappa}{8\pi} \int dtd\varphi\, f^\pm \delta u_\pm
\end{equation}
where $u_\pm$ encode boundary stress-tensor expectation values. Their dynamical nature necessitates introducing compensating boundary terms:
\begin{equation}\label{Binfty}
B_\infty^\pm = \mp \frac{\kappa}{8\pi} \int dtd\varphi\, f^\pm u_\pm
\end{equation}
and at the same time fixing $f^\pm$. In other words, we impose Dirichlet boundary conditions, which fix the asymptotic 2D matric \eqref{2dcurved}. This constrains residual diffeomorphisms to parameters $\epsilon^\pm$ satisfying:
\begin{equation}\label{diffeoConstraint}
\delta_\epsilon f^\pm = \pm \dot{\epsilon}^\pm + \epsilon^\pm f^{\pm\prime} - f^\pm \epsilon^{\pm\prime} = 0.
\end{equation}
General solutions take the form:
\begin{equation}\label{tphi}
\epsilon^\pm = f(\varphi)g_\pm(t \pm \tilde{\varphi}), \quad \tilde{\varphi} = \int^\varphi \frac{d\varphi'}{f(\varphi')}\, ,
\end{equation}
where $g_\pm$ represent arbitrary functions. The Regge-Teitelboim formalism \cite{Regge:1974zd} yields canonical generators:
\begin{equation}
Q^\pm[\epsilon^\pm] = -\frac{\kappa}{8\pi} \oint d\varphi\, \epsilon^\pm u_\pm
\end{equation}
In particular, the Hamiltonian which is associated with the time translation is given by:
\begin{equation}\label{bdyaction}
H = \frac{\kappa}{8\pi} \oint d\varphi\, f(\varphi)(u_+ + u_-),
\end{equation} 
which matches \eqref{inhomo}.
Furthermore, we can define the deformed Virasoro generators as
\be 
\mathcal{L}_n=\oint \frac{d\varphi}{2\pi}f(\varphi) e^{\im n\tilde{\varphi}}T(\varphi).
\ee  
This reproduces the special results of the deformed quantizations proposed in \cite{Ishibashi:2016bey,Okunishi:2016zat} for the SSD and \Mobius deformed Hamiltonians. Our holographic analysis is more general and suitable for all envelope functions. 

\bigskip

Instead of strictly imposing Dirichlet boundary conditions, we can adopt a more general approach by recognizing that the integrability condition leads to the relationship:
\begin{equation}
f^\pm = \frac{\delta H^\pm}{\delta u_\pm}, \label{def}
\end{equation}
where the functional derivative indicates how the Hamiltonian $H^\pm$ depends on the field variables $u_\pm$. Accordingly, we can define the boundary term as:
\begin{equation}
B_\infty^\pm = \mp \frac{\kappa}{8\pi} \int dt\, H^\pm,
\end{equation}
with the Hamiltonian expressed in terms of the phase space variables:
\begin{equation}
H^\pm = \int d\varphi\, \mathcal{H}^\pm(u_\pm, u_\pm', \dots). \label{generalham}
\end{equation}
Choosing different forms of $H^\pm$ corresponds to selecting various boundary conditions, each leading to distinct asymptotic symmetries. In this framework, deriving explicit expressions for the generators of these asymptotic symmetries typically involves solving complex partial differential equations (PDEs). 
To get analytic control, \cite{Perez:2016vqo,Ojeda:2019xih} considers specific integral boundary conditions under which the associated PDEs become integrable. This integrability allows for the construction of a complete set of solutions, providing a clearer understanding of the system. Notably, black hole solutions with non-trivial KdV charges are constructed in \cite{Dymarsky:2020tjh}, with further exploration of their detailed properties in \cite{Chen:2024ysb,Chen:2024lji}. The study of the corresponding inhomogeneous Hamiltonians presents a compelling avenue for exploration, as these Hamiltonians extend beyond previous formulations, which primarily involve linear combinations of Virasoro generators. In this work, we will investigate some holographic properties of these inhomogeneous Hamiltonians in Section \ref{sec:kdv}, while a comprehensive exploration of these deformations is reserved for future research.

\section{General properties of the geometries}
 \label{sec:general}
In this section, we study some general properties of the geometries \eqref{family}. 

\subsection{A generalized Roberts mapping}
First, we notice that introducing the new angular variable \eqref{tphi} along with light-cone coordinates $w=\tilde{\varphi}-t$ and $\bar{w}=\tilde{\varphi}+t$ transforms the metric \eqref{family} into the standard form \eqref{general}. Through the mapping \eqref{mapping}, this further reduces to the \Poincare metric \eqref{poincare}. In the special case where the orbit constant $\fd$ from \eqref{orbitconst} vanishes, we find $\mathcal{L}=\bar{\mathcal{L}}=0$ in \eqref{llbar}, simplifying \eqref{mapping} to:
\begin{align}
\eta &= \frac{4 z f(\varphi)}{z^2 f'(\varphi)^2 + 4 f(\varphi)^2}, \nonumber \\
y &= -t + \tilde{\varphi} + \frac{2 z^2 f'(\varphi)}{z^2 f'(\varphi)^2 + 4 f(\varphi)^2}, \quad \bar{y} = t + \tilde{\varphi} + \frac{2 z^2 f'(\varphi)}{z^2 f'(\varphi)^2 + 4 f(\varphi)^2}, \nonumber \\
t_p &\equiv \frac{\bar{y}-y}{2} = t, \quad 
\varphi_p \equiv \frac{\bar{y}+y}{2} = \tilde{\varphi} + \frac{2 z^2 f'(\varphi)}{z^2 f'(\varphi)^2 + 4 f(\varphi)^2}. \label{phi_p_def}
\end{align}
This mapping reveals that when the angular coordinate $\tilde{\varphi}$ is compact (as in \Mobius- deformations), the original geometry \eqref{family} maps to a proper subregion of \Poincare AdS space. In the case of a flat asymptotic boundary metric, the mapping \eqref{mapping} simplifies to Roberts mapping. Additionally, under the condition that $\mathcal{L}=\bar{\mathcal{L}}=0$, it corresponds to the transformation described in \cite{Skenderis:1999nb}.

The compact spatial identification $\varphi \sim \varphi + 2\pi$ introduces a conical singularity at the locus where the angular metric component vanishes ($g_{\varphi\varphi}=0$):
\begin{equation}
z_s^{-2} = -\frac{u}{4},
\end{equation}
when $u<0$. This singular surface, conventionally termed the ``wall", necessitates surgical excision of its interior region when computing the on-shell action within the Euclidean continuation of the geometry. The removal of the wall-bounded submanifold constitutes a standard regularization procedure to ensure finite gravitational path integral contributions. As a pedagogical example, consider that there exists a coordinate transformation that brings $u$ to a negative constant $u_0<0$. The invariant condition \eqref{orbitconst} forces $f$ to similarly become a constant: $f \to f_0$. The metric consequently simplifies to:
\begin{equation}
ds^2 = \frac{dz^2}{z^2} + \left(\frac{1}{z} - \frac{z u_0}{4}\right)^2 f_0^2 d\tau^2 + \left(\frac{1}{z} + \frac{z u_0}{4}\right)^2 d\varphi^2, \quad \tau = \im t,
\end{equation}
which is degenerate at $z = 2/\sqrt{-u_0}$. Through the coordinate transformation:
\begin{equation}
r = \frac{1}{z} + \frac{z u_0}{4}, \quad \tau_0 = f_0 \tau,
\end{equation}
the metric adopts the manifestly singular form:
\begin{equation}
ds^2 = \frac{dr^2}{r^2 - u_0} + (r^2 - u_0)d\tau_0^2 + r^2 d\varphi^2. \label{constmetric}
\end{equation}
The geometry exhibits a conical singularity at $r=0$ characterized by a deficit angle $2\pi(1-\sqrt{-u_0})$ when $u_0 \neq -1$, while the physical requirement $r \geq 0$ restricts the $z$-coordinate to the interval $0 < z < 2/\sqrt{-u_0}$. This upper bound $ z_s \equiv 2/\sqrt{-u_0}$ precisely corresponds to the location of the wall in the original coordinate system. The critical parameter choice $u_0 = -1$ yields a smooth geometry where the conical singularity vanishes completely. This corresponds to global $\text{AdS}_3$ spacetime, which is holographically dual to the vacuum state of the CFT.

The generalized Roberts mapping \eqref{mapping} establishes a robust framework for investigating holography of the inhomogeneously deformed CFTs. The general strategy is the following:
\begin{itemize}
    \item \textbf{Deformation function selection:} Choose the deformation function $f(\varphi)$.
    
    \item \textbf{Stabilizer equation solution:} Solve the differential constraint \eqref{stabEqn} to determine $u(\varphi)$. The solution space exhibits linear superposition freedom: if $u_*(\varphi)$ satisfies the equation, then $u_*(\varphi) + c_1 f(\varphi)^{-2}$ constitutes a complete solution family, where $c_1$ is a free parameter.
    
    \item \textbf{Orbit invariant computation:} Calculate the orbit invariant quantities $f_0$ and $\fd = u_0 f_0^2$ via \eqref{def0} and \eqref{orbitconst}. Under Dirichlet boundary conditions, $f_0$ becomes uniquely determined by $f(\varphi)$, leaving $\fd$ (or equivalently $u_0$) as the bulk solution parameter.
    
    \item \textbf{Vacuum and black hole specification:} The pure AdS vacuum configuration corresponds to the parameter choice $u_0 = -1$ ($\fd = -f_0^2$). Positive values $\fd > 0$ parameterize smooth black hole geometries when accompanied by appropriate horizon regularity conditions given in \eqref{smooth1} and \eqref{smooth2}.
    
    \item \textbf{Observable calculation:} Transform solutions to the standard coordinate system \eqref{general} to extract the complete Roberts mapping. This enables efficient computation of holographic quantities (e.g., entanglement entropy shown in Section \eqref{section:hee} through:
    \begin{enumerate}
        \item Evaluation in the simplified \Poincare frame;
        \item Inverse mapping application to recover original coordinate dependencies.
    \end{enumerate}
\end{itemize}
\subsection{The black-hole solutions} 
When the conditions
\begin{align}
z_h^2=\frac{4f}{uf-2f''}\geq 0, \label{smooth1}\\
z_h^{-2}\geq z_s^{-2}\geq -\frac{u}{4},\quad\rightarrow \quad 1+\frac{uf}{uf-2f''}>0, \label{smooth2}
\end{align} 
are satisfied, the bulk metric describes a black hole with an event horizon at $z=z_h$. The second condition ensures the black hole horizon hides the $g_{\varphi\varphi}=0$ singularity. The near-horizon geometry after Wick rotation becomes
\begin{align}
ds^2\sim 4{\rho}^2f^2d\tau^2+\frac{4f^2}{uf^2+2{f'}^2-2ff''}d\rho^2+h_{\varphi\varphi}(d\varphi+N^\varphi d\rho)^2
\end{align}
with components:
\begin{align}
h_{\varphi\varphi}&=(uf^2+{f'}^2-2ff'')\left(\frac{uf-f''}{uf^2-2ff''}\right)^2,\\ 
N^\varphi&=\frac{2f'f \sqrt{2f^2-2ff''}}{(uf^2+{f'}^2-2ff'')(uf-f'')}.
\end{align} 
So the condition $h_{\varphi\varphi}>0$, which says the radius of the event horizon should be positive, imposes the constraint
\be 
\fd>0.
\ee 
From the Euclidean smoothness condition, the temperature is found to be
\be 
T=\frac{\sqrt{\fd}}{2\pi}, \label{tem}
\ee 
remarkably maintaining constant despite that horizon position explicitly depends on $\varphi$. The Bekenstein-Hawking entropy, determined through horizon area, takes the explicit form:
\begin{align}
S &= \frac{1}{4G_N}\int_0^{2\pi} d\varphi \sqrt{h_{\varphi\varphi}} \nonumber \\
  &= \frac{\sqrt{\fd}}{4G_N} \int \left|\frac{uf - f''}{uf^2 - 2ff''}\right| d\varphi \label{intS} \\
  &= \frac{\sqrt{\fd}}{4G_N} \int_0^{2\pi} \frac{d\varphi}{|f|}\left(1 + \frac{ff''}{\fd - 2{f'}^2}\right),
\end{align}
where we have assumed that the two smooth conditions \eqref{smooth1} and \eqref{smooth2} are satisfied.
When $f$ has no zeros (\ie is sign-definite), the absolute value $|f|$ can be replaced by $\pm f$. The integral converges, and the second term does not contribute as it constitutes a total derivative. The entropy thus simplifies to
\be 
S=\frac{\sqrt{\fd}}{4G_N}\frac{2\pi}{|f_0|},\quad \frac{1}{f_0}\equiv\frac{1}{2\pi}\int_0^{2\pi}\frac{d\varphi}{f}.\label{def0}
\ee 
This entropy expression depends on two key physical quantities:
\begin{equation}
\text{Entropy density: } \frac{\sqrt{\fd}}{4G_N}, \quad 
\text{Effective system size: } {L}_{{\text{eff}}} \equiv \frac{2\pi}{|f_0|}. 
\end{equation}
It also explicitly demonstrates that the entropy is invariant under Virasoro coadjoint transformations since both $\fd$ (the orbit invariant) and $f_0$ remain constant along any coadjoint orbit. 

\begin{figure}[h]
\begin{center}
	 \includegraphics[scale=0.7]{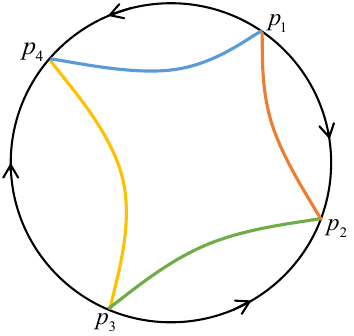}
  \caption{The geometry is partitioned into disconnected horizon segments, illustrated in distinct colors. The orientation of these subregions alternates, as indicated by the arrow.} \label{dishorizon}
\end{center}
\end{figure}

When the function $f$ has zeros coinciding with the poles of the integral \eqref{intS}, the integral becomes formally divergent. From the expression \eqref{smooth1} describing the black hole horizon, we observe that these zeros correspond to locations where the horizon intersects the AdS boundary. Consequently, the zeros of 
$f$ partition the black hole horizon into multiple disconnected segments, as shown in Figure \eqref{dishorizon}. 
The area of each horizon segment is divergent. A similar phenomenon also occurs in the case of the 3D accelerating black hole geometry \cite{Astorino:2011mw,Xu:2011vp,Arenas-Henriquez:2022www} and spacetime banana geometry \cite{Abajian:2023jye,Tian:2024fmo}. One possible way to cure this divergence is to cut out each segment with EOW branes and sew these segments together along the EOW branes, subject to the Israel junction conditions \cite{Israel:1966rt}. This strategy has been used to construct a 3D accelerating black hole with a finite black hole entropy \cite{Arenas-Henriquez:2022www}. However, the price to pay is that the EOW brane also contributes to the black hole entropy through its boundary entropy, rendering an exotic black hole thermodynamics \cite{Tian:2023ine,Tian:2024mew}. In the current situation, we find this approach unsatisfactory because the black hole defined in this way fails to be orbit invariant.

With the requirement of orbit invariance, we propose an alternative regularization and interpretation for bulk solutions when $f$ has simple zeros. By the \Poincare-Hopf theorem, simple zeros of a vector field on $S^1$ necessarily appear in pairs. To demonstrate our proposal, consider a simplified scenario where $f$ has two simple zeros at $\varphi_1,\varphi_2,\varphi_1<\varphi_2$.
Without loss of generality, assume: 
\be 
\begin{cases}
	f<0,\quad \varphi\in(\varphi_1,\varphi_2),\\
	f=0,\quad \varphi=\varphi_1,\varphi_2,\\
	f>0,\quad \text{others},
\end{cases} 
\ee  
with $z_h(\varphi_1)=z_h(\varphi_2)=0$. In regions where $f<0$ and $f>0$, we can introduce new variables:
\be 
d\tilde{\varphi}^+=\frac{d\varphi}{f},\quad d\tilde{\varphi}^-=\frac{d\varphi}{f}.
\ee 
In these coordinates, the $\( f > 0 \)$ region describes a black hole segment, while the $\( f < 0 \)$ region  corresponds to another black hole segment with \textbf{opposite orientation} due to $\( f < 0 \)$. Therefore, the total black hole entropy should be
\be 
S_+(\tilde{\varphi}^+)-S_-(\tilde{\varphi}^-)=\frac{\sqrt{\fd}}{4G_N}\Delta,\quad \Delta\equiv\intbar _0^{2\pi}\frac{1}{f}d\varphi, \label{en}
\ee 
where $\Delta$ is an orbit invariant and finite according to \cite{Witten:1987ty}. In the general case, we always associate a negative entropy (and energy) to the black hole segment with negative $f$. Thus, we can interpret the solution as a non-equilibrium black hole ``emitting" small black holes that carry a fraction of the total energy and entropy of the black hole. It worthes to point out that similar non-equilibrium black holes constructed by gluing thin-shells have been recently studied in \cite{Balasubramanian:2024rek}.

When $f$ has double zeros, the integral \eqref{intS} also diverges. In this case, there is no finite orbit constant, which further suggests that no finite regularized black hole entropy exists in such configurations. The divergence of the entropy directly stems from the infinite effective spatial size of the system. A canonical illustration of this phenomenon is the flat BTZ black hole geometry (often termed the ``BTZ black hole ring'').

\subsection{The holographic mass and the first law}
 Using the standard holographic renormalization dictionary, we compute the energy-momentum tensor associated with the bulk solution. The holographic mass is obtained by integrating the $T_{tt}$ component over the $S^1$ : 
\be \label{eq:adm}
M_{\text{holo}}[\varphi]=\frac{c}{12}\int_0^{2\pi} \frac{d\varphi}{2\pi}\,  fu,
\ee 
where the Brown-Henneaux relation $c=3/(2G_N)$ \cite{Brown:1986nw} has been used. It should be pointed out that this holographic mass is not always equal to the deformed Hamiltonian, as the definition of the energy depends on the choices of asymptotic boundary conditions and the time coordinate. For the Dirichlet boundary condition, the holographic mass coincides with the deformed Hamiltonian and matches the boundary action \eqref{bdyaction}. In contrast, for the general boundary conditions discussed in Sec.~\ref{sec:boundary}, these quantities generally differ, as seen by comparing \eqref{bdyaction} and \eqref{generalham}. Using the identity \eqref{orbitconst} and dropping the total derivative term, the holographic mass \eqref{eq:adm} can be rewritten as
\be
M_{\text{holo}}[\varphi] = \frac{c}{12} \int_0^{2\pi} \frac{d\varphi}{2\pi} \left( \frac{\fd}{f} - \frac{(f')^2}{f} \right).
\ee
Under the Dirichlet boundary condition with $f$ fixed, the holographic mass only depends on the orbit invariant, so the variation of the holographic mass becomes
\be
\delta M_{\text{holo}}[\varphi] = \frac{c}{12} \delta \fd \int_0^{2\pi} \frac{d\varphi}{2\pi} \frac{1}{f}.
\ee
For black hole solutions, we verify the first law of thermodynamics:
\be
T \delta S = \frac{\sqrt{\fd}}{2\pi} \frac{\delta(\sqrt{\fd})}{4G_N} \int \frac{d\varphi}{f} = \delta M_{\text{holo}}[\varphi].
\ee

\subsection{Euclidean on-shell action}
After defining the holographic mass, black hole temperature, and entropy, we proceed to verify that the black hole free energy $\mathcal{F} = M_{\text{holo}} - TS$ can indeed be derived directly from the regularized Euclidean on-shell action:
\begin{equation}
I_{E} = -\frac{1}{16\pi G_N} \int \sqrt{g}\,(R+2)\,dV - \frac{1}{8\pi G_N} \int_{\partial\mathcal{M}} \sqrt{h}\,(K-1)\,d\Sigma,
\end{equation}
where $R$ denotes the Ricci scalar and $K$ represents the trace of the extrinsic curvature at the AdS boundary. We place the AdS boundary at $z = \Lambda^{-1}$ with $\Lambda \gg 1$. We also assume  that the bulk solution corresponding to a smooth black hole occupies the radial domain $z \in (\Lambda^{-1}, z_h)$. Evaluating the radial integral yields the bulk contribution:
\begin{align}
I_{\text{bulk}} &= -\frac{1}{16\pi G_N} \int \sqrt{g}\,(R+2)\,dV =\frac{1}{4\pi G_N}\int \sqrt{g}\,dV\nonumber \\
&= \frac{1}{4\pi G_N} \frac{1}{16} \int d\tau d\varphi \left(8\Lambda^2 f + 8\log\Lambda f'' - \left[4fu - 4f'' + 8f''\log z_h\right]\right),
\end{align}
while the boundary term contributes:
\begin{equation}
I_{\text{bdy}} = -\frac{1}{8\pi G_N} \int \sqrt{h}\,(K-1)\,d\Sigma = -\frac{1}{8\pi G_N} \int dt dx \left(f\Lambda^2 - \frac{f''}{2}\right).
\end{equation}
Combining these contributions and neglecting total derivative terms, we ultimately obtain:
\begin{align}
I_E &= -\frac{1}{16\pi G_N} \int d\tau d\varphi \left(fu - f''\log z_h^2\right) \nonumber \\
&= -\frac{1}{16\pi G_N} \int d\tau d\varphi \left(fu - 2fu + \frac{2\fd(fu - f'')}{f(fu - 2f'')}\right) \nonumber \\
&= \beta(M_{\text{holo}} - TS) = \beta\mathcal{F},
\end{align}
which establishes the expected thermodynamic relation.

\subsection{Holographic entanglement entropy}
\label{section:hee}
Using the generalized Roberts mapping, we can systematically derive the geodesic profile and compute the holographic entanglement entropy via the Ryu-Takayanagi (RT) formula \cite{Ryu:2006bv,Ryu:2006ef,Hubeny:2007xt}. We start with a particular solution in the family \eqref{family}. Introducing the new spatial coordinate
\be 
\tilde{\varphi}=\int^{\tilde{\varphi}}\frac{d\varphi}{f(\varphi)}
\ee   
and the light-cone coordinates $w=\tilde{\varphi}-t,\, \bar{w}=\tilde{\varphi}+t$, we can recast the metric into the standard form \eqref{general} through the identifications
\be 
e^{\phi}=f^2,\quad \fd=4\mathcal{L}=4\bar{\mathcal{L}}.
\ee 
The \Poincare coordinates $Y$ and $\bar{Y}$ are subsequently determined from the differential equation
\be 
\frac{3\bar{y}''^2-2\bar{y}'\bar{y}'''}{\bar{y}'^2}=\frac{3y''^2-2y'y'''}{y'^2}=\fd.
\ee 
The generic solution takes the form
\be 
y(w)=c_3+c_2 \frac{\tanh(\sqrt{\fd/4}(w-c_1))}{\sqrt{\fd/4}},
\ee 
where $c_1,c_2$ and $c_3$ are free parameters originiating from the isometries of the \Poincare AdS$_3$ spacetime. For simplicity, we can choose
\be
Y=y(w)=\begin{cases}
	\tan\frac{\sqrt{-\fd}w}{2},&\quad \fd<0,\\
	\tanh\frac{\sqrt{\fd}w}{2},& \quad \fd>0,
\end{cases}
\ee 
with an analogous expression for $\bar{Y}$. Having derived the required Roberts mapping, we proceed to calculate the holographic entanglement entropy. For a subsystem $A\in [\varphi_1,\varphi_2]$, we need to compute the length of the boundary-anchored geodesic $\Gamma_A$ connecting the two endpoints:
\be 
(\varphi,z)=(\varphi_1,\epsilon),\quad (\varphi,z)=(\varphi_2,\epsilon).
\ee 
Explicitly, the generalized Roberts mapping transforms $(\varphi,z)$ to \Poincare coordinates via
\bea  
&&\(\frac{Y+\bar{Y}}{2},\eta\)=(\Phi,\eta),\nn
&&\Phi=\frac{2 z^2 y'(w)^2 \left(f'(\varphi ) y'(w)-y''(w)\right)}{z^2 \left(y''(w)-f'(\varphi ) y'(w)\right)^2+4 f(\varphi )^2 y'(w)^2}+y(w)|_{w=\tilde{\varphi}},\\
&&\eta=\frac{4 z f(\varphi ) y'(w)^3}{z^2 \left(y''(w)-f'(\varphi ) y'(w)\right)^2+4 f(\varphi )^2 y'(w)^2}|_{w=\tilde{\varphi}}.
\eea  
Substituting $\varphi$ simplifies the mapping to
\be 
\Phi=Y(\varphi )-\frac{2 z^2 Y'(\varphi )^2 Y''(\varphi )}{z^2 Y''(\varphi )^2+4 Y'(\varphi )^2},\quad \eta=\frac{4 z Y'(\varphi )^3}{z^2 Y''(\varphi )^2+4 Y'(\varphi )^2},\label{srm}
\ee
where $Y(\varphi)\equiv y(\tilde{\varphi}(\varphi))$ and $Y'\equiv \partial_\varphi Y,\ Y''\equiv \partial^2_\varphi Y$. In \Poincare AdS spacetime, spacelike geodesics connecting endpoints $(\Phi_1,\Phi_2)$ form semicircles:
\be 
\(\Phi-\frac{\Phi_1+\Phi_2}{2}\)^2+\eta^2=\(\frac{\Phi_2-\Phi_1}{2}\)^2,
\ee 
whose length is given by 
\be 
\Gamma_A=\log\frac{|\Phi_1-\Phi_2|^2}{\eta_1\eta_2}.
\ee 
Through \eqref{srm}, we can easily derive the geodesic and its length in the original geometry. The geodesic length determination requires only the asymptotic form of \eqref{srm}:
\be 
\Phi=Y(\varphi),\quad \eta=zY'(\varphi)
\ee 
leading to
\be 
\Gamma_A=\log\left[\frac{|Y(\varphi_1)-Y(\varphi_2)|^2}{\epsilon^2 |Y'(\varphi_1)Y'(\varphi_2)|}\right],\quad S_A=\frac{\Gamma_A}{4G_N}=\frac{c}{6}\log\left[\frac{|Y(\varphi_1)-Y(\varphi_2)|^2}{\epsilon^2 |Y'(\varphi_1)Y'(\varphi_2)|}\right].
\ee  
It is also convenient to express it in terms of $\tilde{\varphi}$:
\bea  
S_A &=&\frac{c}{6}\log\left[\frac{|y(\tilde{\varphi}_1)-y(\tilde{\varphi}_2)|^2 }{\epsilon^2 |y'(\tilde{\varphi_1})y'(\tilde{\varphi_2})[f(\varphi_1)f(\varphi_2)]^{-1}|}\right],\\
&=&\begin{cases}
	\frac{c}{3}\log\left|\frac{2\sin\frac{\sqrt{-\fd}(\tilde{\varphi}_2-\tilde{\varphi}_1)}{2}}{\sqrt{\fd}\epsilon\sqrt{[f(\varphi_1)f(\varphi_2)]^{-1}}}\right| ,& \fd<0\, ,\\
	\frac{c}{3}\log\left|\frac{2\sinh\frac{\sqrt{\fd}(\tilde{\varphi}_2-\tilde{\varphi}_1)}{2}}{\sqrt{\fd}\epsilon\sqrt{[f(\varphi_1)f(\varphi_2)]^{-1}}}\right| ,& \fd>0\, . \label{gee}
\end{cases}
\eea

\section{Examples}
\label{sec:examples}
In this section, we study examples in each class defined in \eqref{class} of the orbits. We will choose the Dirichlet boundary condition. Our analysis serves dual purposes: first, to bridge existing theoretical frameworks through explicit case comparisons, and second, to extrapolate new physical insights via systematic deformation analysis of benchmark solutions.

\subsection{$S^1$}
\begin{figure}[h]
\begin{center}
    \includegraphics[scale=0.6]{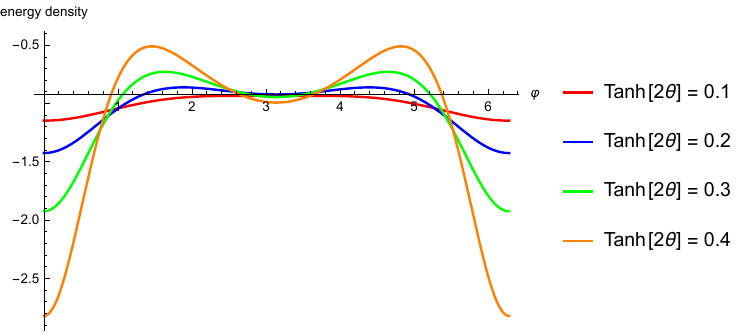}
    \includegraphics[scale=0.6]{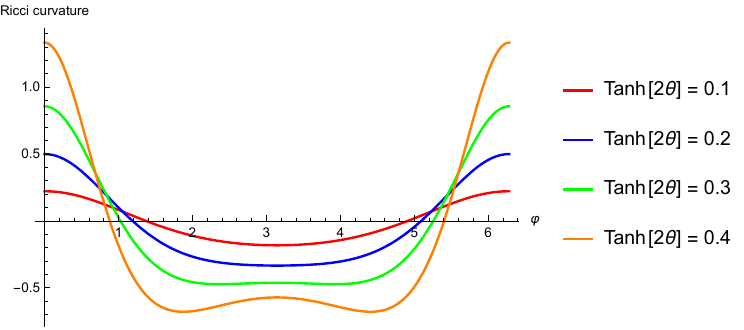}
    \caption{Left: The reduced energy density $\mathcal{H}_r=uf$ of the vacuum state with $f_0=1$ and different values of $\theta$. Right: The Ricci curvature of the 2D curved spacetime with $f_0=1$ and different values of $\theta$.} \label{combined}
\end{center}
\end{figure}

This classification admits further refinement through three distinct subclasses:
\begin{equation}
    \mathcal{C}(\nu), \quad \mathcal{B}_0(b), \quad \mathcal{P}_0^+,
\end{equation}
following the terminology in \cite{Balog:1997zz}. The subclassification derives from the monodromy matrix analysis of Hill's equation solutions, where the $SL(2,\mathbb{R})$ conjugacy class combined with a discrete winding number $\omega\in\mathbb{Z}$ uniquely characterizes each Virasoro coadjoint orbit. We refer to the original work \cite{Balog:1997zz} or Appendix \ref{appendix:virasoro} for details. The critical property of $f(\varphi)$ in this class is that it is sign-definite. This permits us to find a diffeomorphism \eqref{diff} to bring $f$ and $u$ to constants:
\be
    d{\varphi_0} = \frac{f_0}{f(\varphi)}d\varphi:\quad 
    f \mapsto f_0=\left( \frac{1}{2\pi} \int_0^{2\pi} \frac{d\varphi}{f(\varphi)} \right)^{-1}  ,\quad   u \mapsto u_0 = \frac{\fd}{f_0^2}.
\ee 
The refined classification is determined through the $u_0$ parameter space:
\begin{align}
    u_0 < 0,\ u_0 \neq -n^2\ (n\in\mathbb{Z}^+) &: \mathcal{C}(\nu)\ \text{(elliptic)} \label{class-elliptic} \\
    u_0=-n^2\ (n\in\mathbb{Z}^+) &: PSL(2,\mathbb{R}) \ \text{(degenerate parabolic)}\\
    u_0 = 0 &: \mathcal{P}_0^+\ \text{(non-degenerate parabolic, $\omega=0$)} \label{class-parabolic} \\
    u_0 > 0 &: \mathcal{B}_0(b)\ \text{(hyperbolic, $\omega=0$)} \label{class-hyperbolic}
\end{align}
As a paradigmatic example, consider the envelope function:
\begin{equation}
    f(\varphi) = \frac{f_0}{1 - \tanh(2\theta)\cos\varphi}, \label{eq:envelope}
\end{equation} 
which is the reciprocal of $f_{\text{\Mobius}}$. It should be pointed out that s certain aspects of the inhomogeneous Hamiltonian associated with this envelope function were previously investigated in \cite{Lapierre:2020ftq} within Floquet system contexts. Solving the stabilizer equation \eqref{stabEqn} yields:
\be 
u(\varphi)=\frac{u_0f_0^2}{f^2}-\frac{3 f^2 \text{sech}^2(2 \theta )}{{f_0}^2}+\frac{4 f}{{f_0}}-1.
\ee 
Thus, all static solutions are parameterized by the constant $u_0$. Let us first consider the vacuum state with $u_0 = -1$. The corresponding holographic mass is given by:
\begin{align}
    M_{\text{holo,vac}}[\varphi] &= -\frac{c}{12}f_0\left(1 + \frac{1}{2}\sinh^2(2\theta)\cosh(2\theta)\right), \label{vacE}
\end{align}
where we observe that the first term represents the vacuum energy of a 2D CFT defined on a circle of length $L_{\text{eff}} = \frac{2\pi}{f_0}$. The second term, which depends on the parameter $\theta$, arises from the inhomogeneous deformation of the system. Figure~\eqref{combined} demonstrates the vacuum energy density distribution and the Ricci curvature profile of the boundary spacetime. As the parameter $\theta$ increases, the energy density becomes increasingly localized at the point $\varphi = 0$, where the curvature reaches its extrema.

\begin{figure}[htbp]
\centering
\includegraphics[scale=0.7]{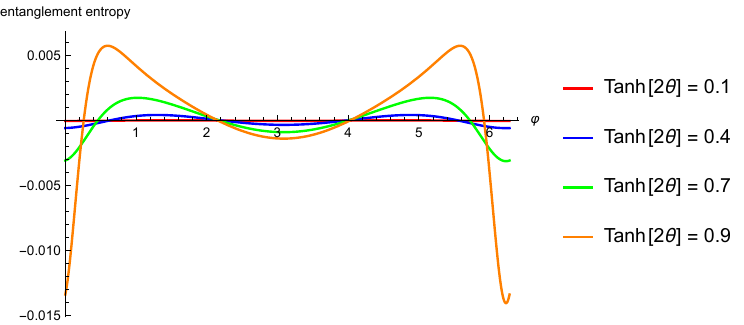}
\caption{Entanglement entropy difference $\Delta S_A = S_A^{\theta} - S_A^{\theta=0}$ for a small interval with $\varphi_2 - \varphi_1 = 0.1$ and $f_0=1$. The unit of the $y$ axis is $\frac{c}{3}$.}
\label{ee1}
\end{figure}

It is straightforward to compute the holographic entanglement entropy for a subsystem consisting of one interval $A\in[\varphi_1,\varphi_2]$:
\be 
S_A=\frac{c}{6}\Gamma_A=\frac{c}{3}\log\frac{\frac{2}{f_0}\sin(\frac{f_0(\tilde{\varphi}_2-\tilde{\varphi}_2)}{2})}{\sqrt{(f(\varphi_1)f(\varphi_2))^{-1}}\epsilon},\quad \tilde{\varphi}=\int_0^\varphi\frac{d\varphi}{f}=\frac{\varphi-\tanh(2\theta)\sin\varphi}{f_0},
\ee 
where $\epsilon$ denotes the UV cut-off, and $\Gamma_A$ corresponds to the length of the geodesic connecting the two endpoints of the integral. This result reveals two distinct effects of the inhomogeneous deformation: 1) modifying the UV cut-off and 2) introducing the inhomogeneity $\varphi\rightarrow \tilde{\varphi}$. Compared to conventional holographic approaches (e.g., \cite{Bai:2024azk}), our method leveraging the generalized Roberts mapping in~\eqref{mapping} provides direct access to the geodesic profile beyond merely determining its length. To study the effect of the inhomogeneity on the entanglement entropy, we fix the interval to be small and plot the entropy difference between the deformed and undeformed theories. As shown in Figure~\eqref{ee1}, the entanglement entropy becomes increasingly concentrated at $\varphi = 0$ as $\theta$ increases.

\begin{figure}[h]
\begin{center}
	 \includegraphics[scale=0.65]{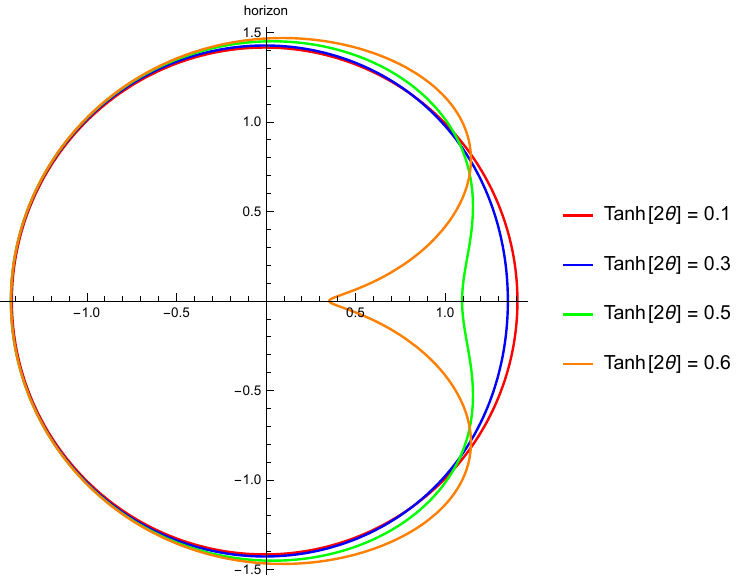}
  \caption{The profile of the black hole horizon: $z_h \text{ vs. } \varphi$ in the polar coordinate. $u_0=10$.} \label{horizon1}
\end{center}
\end{figure}

When $u_0>0$, the solutions are in the black-hole phase. The smoothness conditions \eqref{smooth1} and \eqref{smooth2} require
\begin{equation}
u_{0,\mathrm{BH}} > e^{6\theta} \sinh(2\theta) \cosh^2(2\theta).
\label{bh-condition}
\end{equation}
The horizon geometry is shown in Figure \eqref{horizon1}. The horizon develops a dip at the point $\varphi=0$ and makes contact with the AdS boundary when the deformation parameter $\theta$ is large enough such that the smoothness conditions are violated. 
The black hole free energy is given by
\begin{align}
\mathcal{F} &= M_{\mathrm{holo,BH}} - TS, \nonumber \\
&= \frac{c}{12}f_0\left(u_{0,\mathrm{BH}} - \frac{1}{2}\sinh^2(2\theta)\cosh(2\theta)\right) 
   - \frac{f_0\sqrt{u_{0,\mathrm{BH}}}}{2\pi} \cdot \frac{cf_0\sqrt{u_{0,\mathrm{BH}}}}{6} \cdot \frac{2\pi}{f_0}, \nonumber \\
&= -\frac{c}{12}f_0\left(u_{0,\mathrm{BH}} + \frac{1}{2}\sinh^2(2\theta)\cosh(2\theta)\right).
\label{bh-free-energy}
\end{align}
Comparing with the vacuum energy~\eqref{vacE}, the Hawking-Page transition occurs when
\begin{equation}
\mathcal{F} = M_{\mathrm{holo,vac}}[\varphi], \quad \Rightarrow \quad u_{0,\mathrm{BH}} = 1, \quad \Rightarrow \quad \left(\frac{2\pi T}{f_0}\right)^2 = (TL_{\mathrm{eff}})^2 = 1.
\label{hp-condition}
\end{equation}The holographic entanglement entropy calculation in the black hole geometry follows analogous procedures to the vacuum case. We therefore omit the details here.

\subsection{$PSL(2,\mathbb{R})$}
\begin{figure}[h]
\begin{center}
    \includegraphics[scale=0.6]{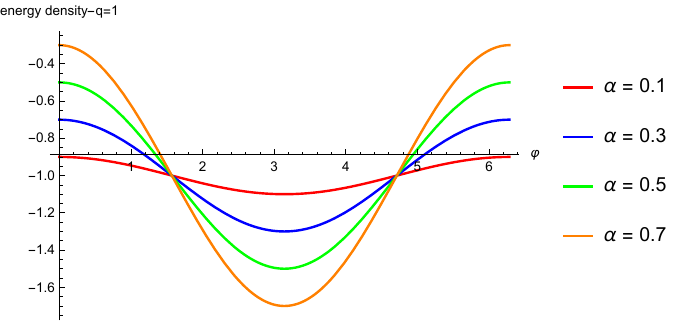}
    \includegraphics[scale=0.6]{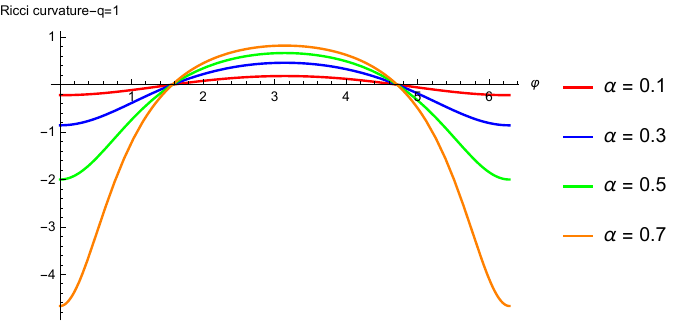}
    \includegraphics[scale=0.6]{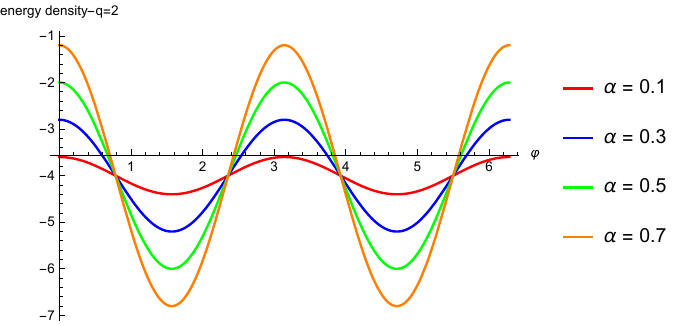}
    \includegraphics[scale=0.6]{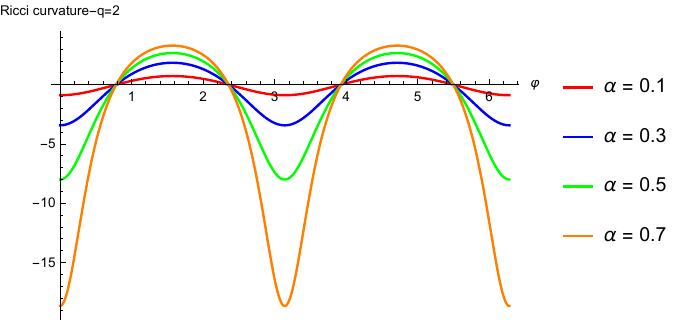}
    \caption{Left: The reduced energy density $\mathcal{H}_r = uf$ of the vacuum state for different values of $\alpha$ and $q$. Right: The Ricci curvature of the 2D curved spacetime for varying values of $\alpha$ and $q$.}
    \label{combined2}
\end{center}
\end{figure}

\begin{figure}[htbp]
\centering
\includegraphics[scale=1]{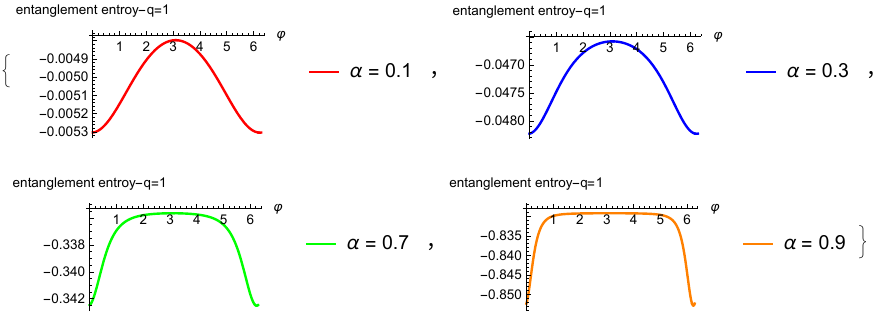}
\includegraphics[scale=1]{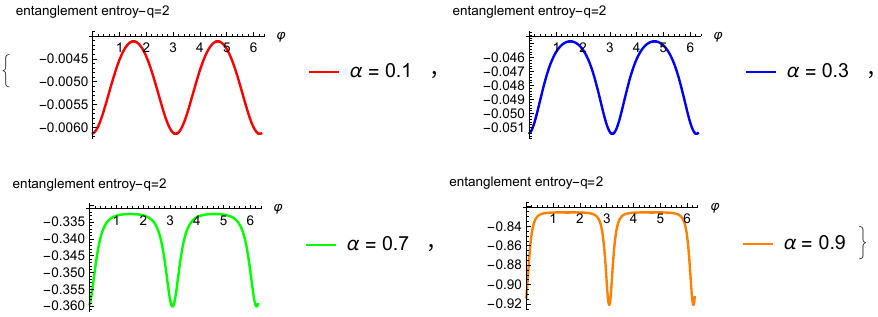}
\caption{Entanglement entropy difference $\Delta S_A = S_A^{\alpha} - S_A^{\alpha=0}$ for a small interval with $\varphi_2 - \varphi_1 = 0.1$ and $q=1,2$. The unit of the $y$ axis is $\frac{c}{3}$.}
\label{ee2}
\end{figure}

The q-\Mobius deformation is in this class. In this section, we analyze this deformation through our holographic framework.
Let us consider the envelope function $f(\varphi)=1-\alpha \cos(q\varphi),\, \alpha\equiv\tanh(2\theta)$. Solving the stabilizer equation \eqref{stabEqn}, we find that 
\be 
u(\varphi)=-q^2+\frac{q^2(1+\gamma) }{f^2(\varphi)},\label{Mobiusu}
\ee 
which depends on a free parameter $\gamma$. Since $f(\varphi)$ is positive-definite, we can introduce the diffeomorphism \eqref{diff} such that
\begin{align}
\varphi &\mapsto {\varphi_0} = 2\arctan\left( \sqrt{\frac{1+\alpha}{1-\alpha}}\tan\left(\frac{q\varphi}{2}\right) \right) \label{m1} \\
f(\varphi) &\mapsto f_0 = q\sqrt{1-\alpha^2} \nonumber \\
u(\varphi) &\mapsto u_0 = \frac{\alpha^2 + \gamma}{1-\alpha^2} ,\quad \fd=q^2(\alpha^2+\gamma).\nonumber
\end{align}
The mapping \eqref{m1} is one-to-one only when $q=1$. For values of $q >1$, the mapping loses its injectivity due to periodic overlaps. For the vacuum state with $u_0=\gamma=-1$, the holographic mass is given by:
\be 
M_{\text{holo,vac}}[\varphi]=-\frac{c}{12}q^2.
\ee 
The energy density and Ricci curvature of the 2D boundary metric are given by:
\begin{equation}
\mathcal{H} = -\frac{c}{12}q^2(1-\alpha\cos{q\varphi}), \quad R = \frac{2 \alpha q^2 \cos (q \varphi)}{\alpha \cos (q \varphi) - 1}.
\end{equation}
As illustrated in figure \ref{combined2}, the relationship between the energy density and the Ricci curvature resembles the previous case: 1) The energy density profile appears as an inverted version of the Ricci curvature profile; 2) As the deformation parameter $\alpha$ increases, the energy density becomes increasingly concentrated at the locations of the extrema of the Ricci curvature.
The holographic entanglement entropy of the subsystem $A\in[\varphi_1,\varphi_2]$ can be determined in a similar manner:
\be 
S_A=\frac{c}{6}\Gamma_A=\frac{c}{3}\log\frac{\frac{2}{f_0}\sin(\frac{f_0(\tilde{\varphi}_2-\tilde{\varphi}_2)}{2})}{\sqrt{(f(\varphi_1)f(\varphi_2))^{-1}}\epsilon},\quad  \tilde{\varphi}=\frac{\varphi_0}{f_0},
\ee 
where ${\varphi}_0$ is defined in equation \eqref{m1}. This expression is consistent with the one derived from the field theory perspective in \cite{Miyata:2024gvr}. For a short interval, the inhomogeneity of the entanglement entropy is illustrated in Figure \eqref{ee2}. As the parameter $\alpha$ approaches $1$, the entanglement entropy also becomes increasingly concentrated at the locations of the extrema of the Ricci curvature. In the limit of $\alpha=1$, these locations coincide with the zeros of the envelope function $f$.

\begin{figure}[htbp]
\centering
\includegraphics[scale=0.55]{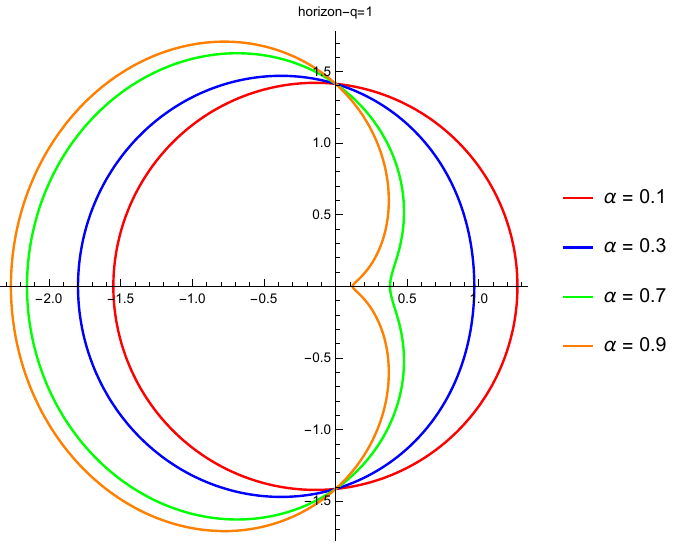}
\includegraphics[scale=0.55]{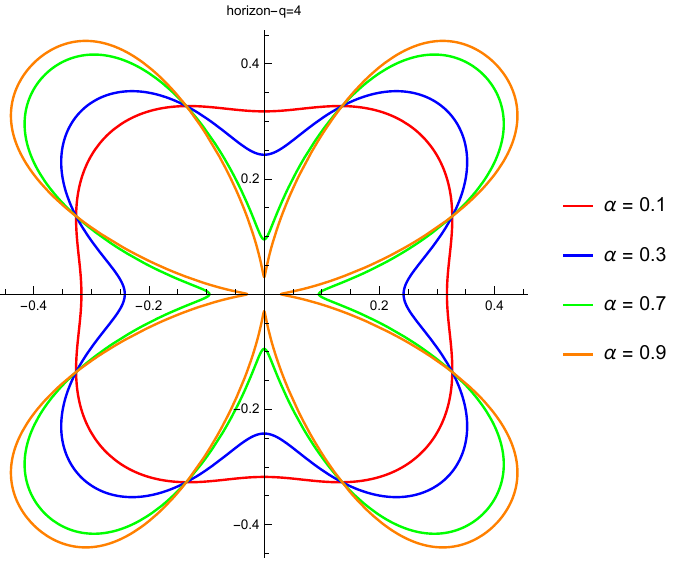}
\caption{The profiles of the black hole horizon: $z_h \text{ vs. } \varphi$ in the polar parametrization. $\gamma=2$. As $\alpha \to 1^-$, the horizon approaches the AdS conformal boundary near the would-be zeros of $f$.}
\label{h2}
\end{figure}

When $u_0>0$, the geometries are in the black-hole phase. We find that the smoothness conditions can be satisfied when
\be 
\gamma>\alpha.
\ee 
Interestingly, the smoothness conditions impose a lower bound on the black hole mass:
\begin{equation}
  M_{\text{holo,BH}} \geq \frac{c}{12} \left[ -q^2 + q \sqrt{\frac{1 + \alpha}{1 - \alpha}} \right].
\end{equation}
The horizon of the black hole is located at
\be 
z_h^{2}=\frac{4 (1-\alpha  \cos (q \varphi))^2}{q^2 \left(\gamma +\alpha ^2 \cos ^2(q \varphi)\right)},
\ee 
with representative examples shown in Figure \eqref{h2}.
The black hole free energy is given by
\begin{align}
  \mathcal{F} &= M_{\text{holo,BH}} - TS \nonumber \\
  &= \frac{c}{12} q \left( -q + \frac{1 + \gamma}{\sqrt{1 - \alpha^2}} \right) - \frac{q \sqrt{\alpha^2 + \gamma}}{2\pi} \cdot \frac{c q \sqrt{\alpha^2 + \gamma}}{6} \cdot \frac{2\pi}{q \sqrt{1 - \alpha^2}} \nonumber \\
  &= -\frac{c}{12} q^2 + \frac{c}{12} \frac{q(1 - 2\alpha^2 - \gamma)}{\sqrt{1 - \alpha^2}}.
\end{align}
Comparing with $ M_{\text{holo,vac}}[\varphi]$, we identify the Hawking-Page transition condition:
\begin{equation}
  \gamma = 1 - 2\alpha^2 \quad \Rightarrow \quad (TL_{\text{eff}})^2 = 1.
\end{equation}
This demonstrates that the Hawking-Page temperature remains invariant under \Mobius deformation, consistent with the findings in \cite{Miyata:2024gvr} through alternative entropy arguments. Remarkably, our analysis establishes that the Hawking-Page temperature persists unchanged under \textit{any} deformation preserving the sign-definiteness of $f(\varphi)$. This invariance stems from the deformation affecting energy levels identically in both black hole and non-black hole phases while leaving black hole entropy and temperature unmodified. 

Similarly, we find the holographic entropy of the subsystem $A$ is given by \eqref{gee}. Here, we assume the interval is short so that only the geodesic connecting the two endpoints contributes. This expression agrees with the field theory derivation in \cite{Miyata:2024gvr}. Note that the relationship between the effective coordinate $\tilde{\varphi}/f_0$ and $\varphi$ can alternatively be expressed as
\begin{equation}
e^{i\tilde{\varphi}/f_0} = \left(\frac{\sinh\theta - e^{iq\varphi}\cosh\theta}{\sinh\theta e^{iq\varphi} - \cosh\theta}\right)^{\frac{1}{q\sqrt{1 - \tanh(2\theta)}}},
\end{equation}
which does not constitute an exact ${PSL}^{(n)}(2,\mathbb{R})$ transformation due to improper normalization of $f(\varphi)$. The correctly normalized one should be \cite{Okunishi:2016zat}:
\begin{equation}
f(\varphi) = \frac{1}{\cosh(2\theta)}\left[1 - \tanh(2\theta)\cos(q\varphi)\right].
\end{equation}
Having reproduced the main results from \cite{Miyata:2024gvr} in our holographic framework, we now present a novel geometric interpretation of the deformation. The key observation is that the vacuum state geometry ($u_0 = -1$) of the $q=2$ \Mobius deformation with the envelope function:
\begin{equation}
f(\varphi) = 1 - a^2\sin^2\varphi, \label{cf}
\end{equation}
corresponds to a distinctive subclass of 3D C-metrics \cite{Astorino:2011mw,Xu:2011vp,Arenas-Henriquez:2022www}\footnote{also see \cite{Arenas-Henriquez:2023hur,Kubiznak:2024ijq,Fontana:2024odl,Tian:2024mew,Cisterna:2023qhh}.}. The 3D C-metrics are a family of AdS solutions describing an accelerating particle or black hole. The metric of the 3D C-metric takes the general form:
\begin{align}
ds^2 &= \frac{1}{\Omega^2}\left[-P(y)a^2dt^2 + \frac{dy^2}{P(y)} + \frac{dx^2}{Q(x)}\right], \\
\Omega &= a(x - y),
\end{align}
classified into three types:
\begin{equation} \label{eq:3class}
\begin{array}{|c|c|c|c|}
\hline
\text{Class} & Q(x) & P(y) & \text{Maximal range of } x \\
\hline
I & 1 - x^2 & \frac{1}{a^2} + (y^2 - 1) & |x| < 1 \\
II & x^2 - 1 & \frac{1}{a^2} + (1 - y^2) & x < -1 \text{ or } x > 1 \\
III & 1 + x^2 & \frac{1}{a^2} - (1 + y^2) & \mathbb{R} \\
\hline
\end{array}
\end{equation}
where $a$ parameterizes acceleration. Focusing on Type I, the FG gauge of the metric is:
\begin{equation}
ds^2 = \frac{dz^2}{z^2} + \frac{g^{(0)}_{ij}dx^idx^j}{z^2} + g^{(2)}_{ij}dx^idx^j + z^2 g^{(4)}_{ij}dx^idx^j
\end{equation}
with components:
\begin{align}
g^{(0)}_{ij}dx^idx^j &= \omega(\xi)^2\left(-dt^2 + \frac{d\xi^2}{(1 - \xi^2)\Gamma(\xi)^2}\right), \quad \Gamma = 1 - a^2 + a^2\xi^2, \label{eq:gamma} \\
g_{tt}^{(2)} &= -\frac{1 - a^2}{2} - \frac{(1 - \xi^2)\Gamma^2\omega'^2}{2\omega^2}, \nonumber \\
g_{\xi\xi}^{(2)} &= -\frac{1 - a^2}{2(1 - \xi^2)\Gamma^2} - \frac{(1 - 3a^2(1 - \xi^2))\xi \omega'}{\Gamma(1 - \xi^2)\omega} + \frac{2\omega\omega'' - 3\omega'^2}{2\omega^2}, \nonumber \\
g^{(4)} &= \frac{1}{4}g^{(2)}{g^{(0)}}^{-1}g^{(2)}. \nonumber
\end{align}
The reality condition $\Gamma > 0$ from \eqref{eq:gamma} implies:
\begin{equation}
\xi^2 > \frac{a^2 - 1}{a^2},
\end{equation}
automatically satisfied in the slow phase ($a < 1$) but constraining the rapid phase ($a > 1$). For a non-trivial conformal factor, the boundary metric $g^{(0)}$ is not flat. To relate this C-metric to our current discussion, we choose the conformal factor as follows:
\begin{equation}
\omega^2 = \Gamma(\xi)^2, \label{conformalfactor}
\end{equation}
which allows the boundary metric to be expressed as:
\begin{equation}
ds^2 = -(1 - a^2\sin^2(\varphi))^2 dt^2 + d\varphi^2, \quad \text{where } \xi \equiv \cos(\varphi),
\end{equation}
which exactly corresponds to \eqref{cf}. Additionally, from the FG gauge of the C-metric, we can identify the corresponding orbit constant as:
\begin{equation}
\fd= a^2 - 1.
\end{equation}
Thus, we can equate the acceleration $a$ with the deformation parameter $a$ in \eqref{cf}. The geometry of the C-metric becomes simpler in the spherical coordinates:
\begin{align} \label{eq:class1_smetric}
ds^2 &= \frac{1}{(1 + ar\cos\varphi)^2}\left[-f(r)dt^2 + \frac{dr^2}{f(r)} + r^2d\varphi^2\right], \\
f(r) &= 1 + (1 - a^2)r^2.
\end{align}
It should be noted that the transformation between the C-metric coordinates and the FG coordinates was only known in an expansion form. However, with the help of the generalized Roberts mapping, this transformation can now be explicitly derived, even for the special choice of the conformal factor presented in \eqref{conformalfactor}. 
Another interesting aspect of this interpretation is that the metric is not presented in the FG gauge, but rather in more ``physical" coordinates, which offer a clearer physical meaning.

Taking the limit $a \rightarrow 1$, the deformation described by \eqref{cf} becomes the Cosine-Square-Deformation (CSD). Furthermore, if we replace $\varphi $ with $\frac{\pi}{2} - \varphi$, it transforms into the SSD deformation. Thus, we expect that the 3D C-metric in the saturated phase serves as the holographic dual of the vacuum state in either the CSD or SSD deformation. Compared to the metric in the FG gauge, the C-metric offers a more transparent physical interpretation. In the C-metric coordinates, the metric is given by:
\begin{equation}
ds^2 = \frac{1}{(x-y)^2}\left(-y^2 dt^2 + \frac{dy^2}{y^2} + \frac{dx^2}{1-x^2}\right), \quad y = -\frac{1}{r},
\end{equation}
which describes a black hole with a horizon at $y = 0$. 
It is important to note that the horizon touches the AdS boundary at $x - y = 0$, suggesting that, naively, the black hole entropy diverges. It has been proposed in \cite{Arenas-Henriquez:2022www} that we can introduce EOW branes to modify the geometry, resulting in a black hole with finite entropy. This implies that we can add boundaries that satisfy conformal boundary conditions to regularize the CSD or SSD deformation. Similar ideas have been explored in \cite{Bernamonti:2024fgx} in the context of driven CFTs.

\subsection{$\tilde{T}_{n,\pm }$}
\begin{figure}[h]
\begin{center}
    \includegraphics[scale=0.65]{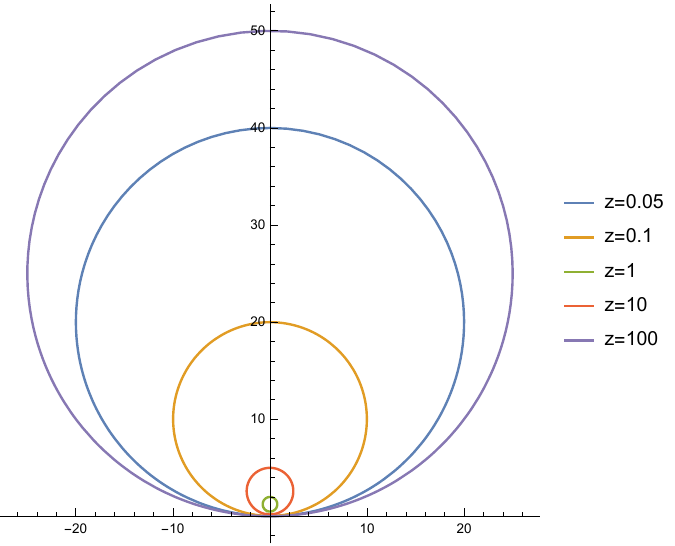}
    \caption{Constant-$z$ surfaces in the \Poincare AdS$_3$ coordinates are circles.} 
    \label{ssdz}
\end{center}
\end{figure}

The stabilizer $f(x)$ has double zeros in this class. However, despite $f_{\text{SSD}}$ having double zeros, it  belongs to the $PSL(2,\mathbb{R})$ class because $f_{\text{SSD}}$ can be rewritten as
\be 
f_{\text{SSD}}=1-2(e^{\im \varphi}+e^{-\im \varphi}),
\ee 
where $\{\partial_\varphi, e^{\pm\im \varphi}\partial_\varphi\}$ form the basis vectors of the $PSL(2,\mathbb{R})$ group. As demonstrated in \cite{Witten:1987ty}, the class $\tilde{T}_{n,\pm}$ emerges as a perturbation of the stabilizer:
\begin{equation}
f_{n\text{-SSD}} = 2\sin^2\left(\frac{n\varphi}{2}\right),
\end{equation}
revealing significant similarities between SSD and $\tilde{T}_{n,\pm}$ deformations. First, we analyze the SSD deformation in more detail. Solving the stabilizer equation with $f=f_{\text{SSD}}$ yields:
\be 
u(\varphi)=-1+\frac{\fd}{4}\csc^4(\frac{\varphi}{2}),
\ee 
which is parameterized by the orbit constant $\fd$ from \eqref{orbitconst}. In particular, when $\fd=0$, the geometry can be transformed into the \Poincare AdS$_3$ through:
\bea 
\eta=\frac{4z}{\left(z^2-4\right) \cos (\varphi)+z^2+4},\quad \varphi_p=\frac{\left(z^2-4\right) \sin (\varphi)}{\left(z^2-4\right) \cos (\varphi)+z^2+4},\quad t_p=t,\label{ssdtr}
\eea 
which satisfy the equation:
\begin{equation}
\left(\eta - \frac{z^2 + 4}{4z}\right)^2 + \varphi_p^2 = \left(\frac{z^2 - 4}{4z}\right)^2.
\end{equation}
This implies that the constant-$z$ surfaces are mapped into circles in the \Poincare AdS$_3$ coordinates as illustrated in Figure \eqref{ssdz}.
In the $z\rightarrow \epsilon$ limit \eqref{ssdtr} reduces to:
\begin{equation}
\eta \sim \frac{\epsilon}{2\sin^2(\varphi/2)}, \quad \varphi_p \sim -\cot\left(\frac{\varphi}{2}\right), \quad t_p = t,
\end{equation}
which demonstrates that the zeros of $f(\varphi)$ correspond to spatial infinities in the \Poincare coordinates. 
The SSD deformed theory has an infinite effective size \cite{Ishibashi:2015jba}. To regularize it, we introduce a cutoff $\delta$ defined as
\be 
L_{\text{eff}}={\int_{0+\delta}^{2\pi-\delta}\frac{d\varphi}{f(\varphi)}}=2\cot(\frac{\delta}{2}).
\ee  
Then we can introduce the diffeomorphism to define the orbit parameters:
\begin{align}
    \varphi &\rightarrow {\varphi}_0 = -f_0 \cot\frac{\varphi}{2}, \\
    f &\rightarrow f_0 = \frac{2\pi}{L_{\text{eff}}}, \\
    u &\rightarrow u_0 = \fd\left(\frac{2\pi}{L_{\text{eff}}}\right)^{-2}.
    \label{eq:diffeo}
\end{align}
The vacuum state corresponds to $u_0=-1$ \ie $\fd=-(2\pi/L_{\text{eff}})^2$ and its holographic mass is 
\be 
M_{\text{holo}}=-\frac{c}{12}+\frac{c}{12}\frac{L_{\text{eff}}}{2\pi}\fd=-\frac{c}{12}-\frac{c}{12}\frac{2\pi}{L_{\text{eff}}},
\ee 
which reduces to the vacuum energy of the undeformed theory as we send the cutoff $\delta$ to zero. Nevertheless, the energy density maintains certain inhomogeneity characterized by:
\be 
\lim_{L_{eff}\rightarrow \infty}\mathcal{H}=\frac{c}{12}\frac{\cos\varphi-1}{2\pi}=-\frac{c}{24\pi}\frac{2}{2-R},\quad R=\frac{2\cos\varphi}{\cos\varphi-2},
\ee 
where $R$ is the Ricci scalar of the 2D boundary spacetime. The holographic entanglement entropy of the subsystem $A$ is given by
\be 
S_A=\frac{c}{6}\Gamma_A=\frac{c}{3}\log\frac{\frac{2}{f_0}\sin(\frac{f_0(\tilde{\varphi}_2-\tilde{\varphi}_2)}{2})}{\sqrt{(f(\varphi_1)f(\varphi_2))^{-1}}\epsilon},\quad  \tilde{\varphi}=\frac{\varphi_0}{f_0},\label{ssdee}
\ee 
where we have assumed that $(f(\varphi_1)f(\varphi_2))^{-1}\epsilon \ll 1$. In other words, we assume that the Roberts mapping should properly transform boundary points between the AdS geometry described by \eqref{family} and the \Poincare coordinate system. The expression \eqref{ssdee} for the entanglement entropy becomes invalid as $\varphi_i$ approach the geometric singular point $\varphi=0$, as it violates our fundamental assumption that $(f(\varphi_1)f(\varphi_2))^{-1}\epsilon \ll 1$.

When $\fd>0$, the solution can also be in the black-hole phase. We find that the smoothness conditions can be satisfied when $\fd>2$. The horizon is located at
\be 
z_h=4\sin^2\frac{\varphi}{2}\sqrt{\frac{2}{2\fd+\cos(2\varphi)-1}},
\ee 
which touches the AdS boundary at the zeros of $f(\varphi)$ as shown in Figure \ref{hssd}.
\begin{figure}[h]
\begin{center}
    \includegraphics[scale=0.7]{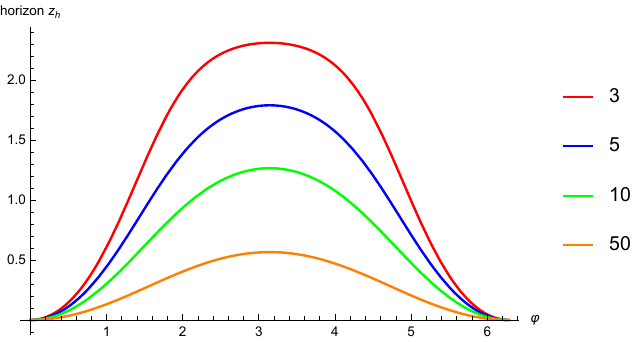}
    \caption{The profiles of the black hole horizon: $z_h \text{ vs. } \varphi$ with different values of $\fd$. The horizon exhibits a self-intersection at $\varphi=0$ on the AdS boundary.} 
    \label{hssd}
\end{center}
\end{figure}
With our regularization, the temperature and the black hole entropy are given by
\bea 
T=\frac{\sqrt{\fd}}{2\pi},\quad S=\frac{c\sqrt{\fd}}{6}L_{\text{eff}}.
\eea 
The holographic entanglement entropy of the subsystem in the black-hole phase is given by
\begin{equation}
S_A = \frac{c}{6}\Gamma_A = \frac{c}{3}\log\left(\frac{2\sinh\left(\frac{\sqrt{\fd}}{2}(\cot{\varphi}_1 - \cot{\varphi}_2)\right)}{\sqrt{\fd}\epsilon\sqrt{(f(\varphi_1)f(\varphi_2))^{-1}}}\right).
\end{equation}

Next, we consider the $\tilde{T}_{n,q}$ deformation defined by the envelope function
\begin{equation}
f_{\tilde{T}_{n,q}}(\varphi) = \frac{\sin^2\left(\frac{n\varphi}{2}\right)}{1 + \frac{q}{2\pi}\sin^2\left(\frac{n\varphi}{2}\right)}.
\end{equation}
The analysis closely resembles the SSD case, so we will omit some details for brevity.
For direct comparison with SSD deformation results, we specialize to the $n=1$ case. Solving the stabilizer equation yields:
\be 
u=\left(q+2 \pi  \csc ^2\left(\frac{\varphi }{2}\right)\right)^2 \left(\frac{\fd}{4 \pi ^2}-\frac{8 \pi  (\cos (\varphi )-1)^2 (q \cos (\varphi )+2 (q+\pi ))}{(q (-\cos (\varphi ))+q+4 \pi )^4}\right)
\ee  
The regularized effective length becomes
\begin{equation}
L_{\text{eff}} = \int_{0+\delta}^{2\pi-\delta} \frac{d\varphi}{f} = q + 4\cot\left(\frac{\delta}{2}\right).
\end{equation} 
In the black hole phase, this leads to the holographic mass and entropy expressions:
\begin{align}
M_{\text{holo}}[\varphi] &= \frac{c}{12}\frac{\fd}{2\pi}q + \frac{c}{12}\frac{\fd}{2\pi}4\cot\left(\frac{\delta}{2}\right), \label{mtn} \\
S &= \frac{c\sqrt{\fd}}{6}q + \frac{c\sqrt{\fd}}{6}4\cot\left(\frac{\delta}{2}\right). \label{stn}
\end{align}
The first terms in \eqref{mtn} and \eqref{stn} are orbit invariants, suggesting they correspond to physically observable quantities. It would be very interesting to explore the $\tilde{T}_{n,q}$ deformation in field theories or lattice models directly.

\subsection{$T_{n,\Delta}$}
\begin{figure}[h]
\begin{center}
	 \includegraphics[scale=0.4]{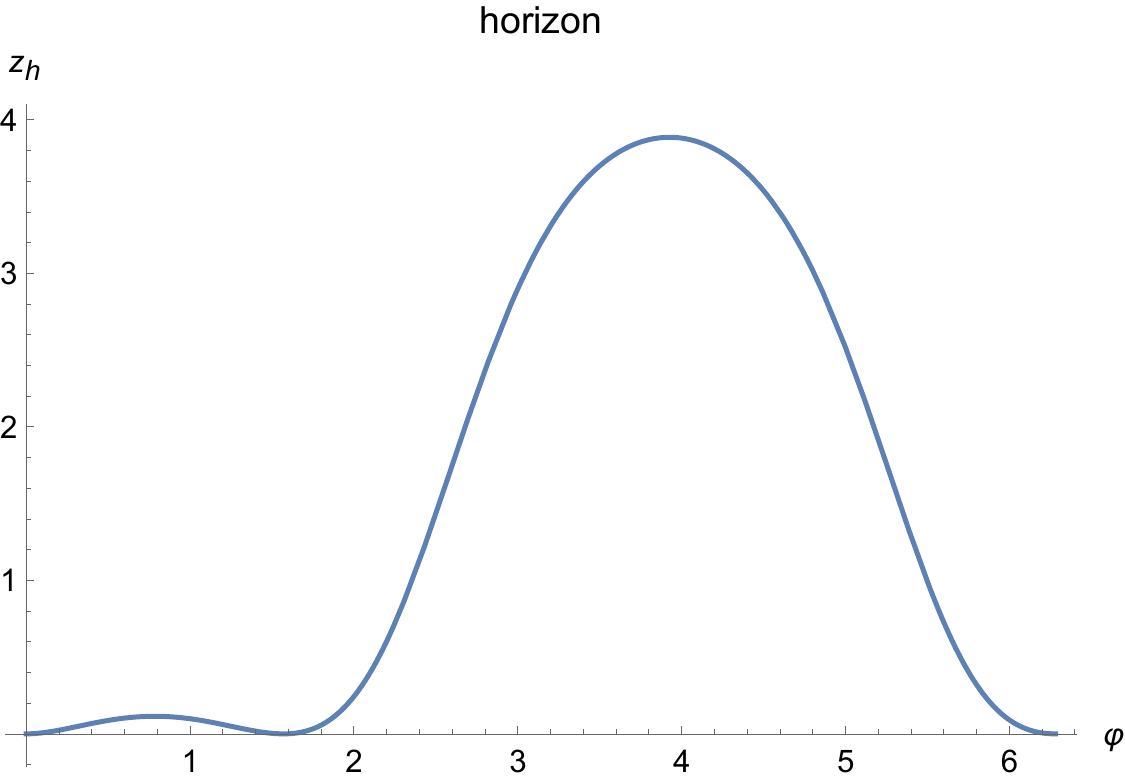}
	 \includegraphics[scale=0.6]{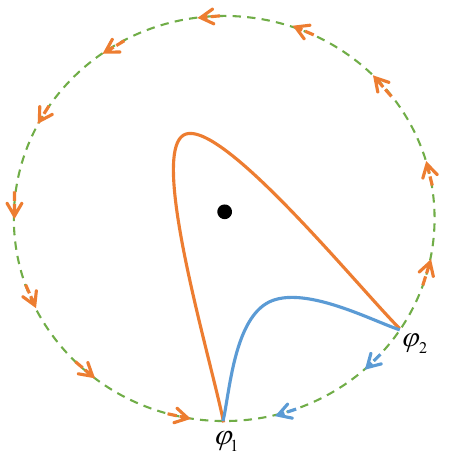}
  \caption{The horizon touches on the AdS boundary at $\varphi=0,\frac{\pi}{2}$ and it consists of two pieces. The right panel shows a schematic representation of the geometry: the dotted line represents the AdS boundary, while the blue and orange curves represent the two segments of the horizon.} \label{h1}
\end{center}
\end{figure}

In this class, the stabilizer has simple zeros. The simplest choice is
\be 
f(\varphi)=k\sin(\frac{n\varphi}{2})\sin(\frac{n(\varphi-\varphi_0)}{2}),\quad 0<\varphi_0<\frac{2\pi}{n}\label{fs0s}
\ee 
whose zeros are given by
\be 
\varphi=\frac{2m\pi }{n},\quad \frac{2m\pi }{n}+\varphi_0,\quad m=0,1,\dots.
\ee 
Despite this construction, the function actually belongs to the $PSL(2,\mathbb{R})$ class, as the corresponding inhomogeneous Hamiltonian can be expressed as a linear combination of $\{L_0, L_n, L_{-n}\}$. A legitimate $T_{n,\Delta}$ orbit can be considered as its perturbation \cite{Witten:1987ty}. So let us study \eqref{fs0s} first. Solving the stabilizer equation \eqref{stabEqn}, we find
\be 
u(\varphi)=\frac{\gamma+n^2\cos^2(\frac{n\varphi_0}{2})}{4 \sin^2(\frac{n\varphi}{2})\sin^2(\frac{n(\varphi-\varphi_0)}{2})}-n^2
\ee  
where $\gamma$ is a free parameter. The orbit constant is given by \footnote{In the literature such as \cite{Witten:1987ty}, it is usually suggested that at zeros of $f(\varphi)$, the invariant reduces to
\be 
f'^2=\fd.
\ee 
However, this holds true only when $u(\varphi)$ is regular at the zeros. In our case, this is not necessarily true. Specifically, we can compute
\be 
f'^2=\frac{1}{4}k^2 n^2\sin^2(\frac{n\varphi_0}{2}). \label{cs2}
\ee 
Indeed, if we choose the parameter $\gamma$ such that \eqref{cs1} and \eqref{cs2} are equal, then $u(\varphi)$ will be regular at the zeros:
\begin{equation}
u(\varphi) \bigg|_{\gamma = -n^2 \cos^2\left(\frac{n\varphi_0}{2}\right)} = -n^2.
\end{equation} 
}
\be 
u f^2+f'^2-2ff''=\fd=\frac{1}{4}k^2(\gamma+n^2).\label{cs1}
\ee 
Notably, this deformation exhibits fundamentally different behavior from the SSD deformation. The effective system size vanishes identically:
\begin{equation}
L_{\text{eff}} = \Delta \equiv \intbar_0^{2\pi} \frac{1}{f} d\varphi = 0. \label{ssdleff}
\end{equation}
Furthermore, the holographic mass is also zero:
\begin{equation}
M_{\text{holo}}[\varphi] = \frac{c}{12} \int_0^{2\pi} u f \, d\varphi = \frac{c}{12} \int_0^{2\pi} \left( \frac{\gamma + n^2\cos^2\left(\frac{n\varphi_0}{2}\right)}{4} - n^2 f^2 \right) d\varphi = 0.
\end{equation}
As discussed in Section \ref{sec:general}, the geometry is segmented by the horizon. Let us illustrate this with our specific example. The black hole smoothness conditions can be satisfied when
\be
\gamma>n^2|\cos\frac{n\varphi_0}{2}|. \label{smooth_cond}
\ee   
The temperature of the black hole is given by
\begin{equation}
T = \frac{k\sqrt{\gamma + n^2}}{4\pi}. \label{temp_eq}
\end{equation} 
Because the effective length is zero, the black hole entropy is also zero. The location of the black hole horizon is determined by
\be 
z_h^2=\frac{32 \sin ^2\left(\frac{n \varphi}{2}\right) \sin ^2\left(\frac{1}{2} n (\varphi-{\varphi_0})\right)}{2 \gamma+n^2 \cos (n (2 \varphi-{\varphi_0}))+n^2}.
\ee 
To visualize this, let us consider a concrete example with $n=k=1,\gamma=2,\varphi_0=\frac{\pi}{2}$. This choice satisfies the smooth condition stated in \eqref{smooth_cond}. The vector field is negative in the domain $(0,\frac{\pi}{2})$. The black hole horizon consists of two distinct pieces, as shown in Figure \eqref{h1}.

To achieve a non-zero effective length and black hole entropy, we consider the $T_{n,\Delta\equiv 4\pi b}$ deformation with the envelope function given by
\begin{align}
f(\varphi) &= \frac{2}{nF}\cos\left(\frac{n\varphi}{2}\right) \left( \sin\left(\frac{n\varphi}{2}\right) + \frac{b}{n}\cos\left(\frac{n\varphi}{2}\right) \right), \label{hyper} \\
F(\varphi) &= \cos^2\left(\frac{n\varphi}{2}\right) + \left( \sin\left(\frac{n\varphi}{2}\right) + \frac{2b}{n}\cos\left(\frac{n\varphi}{2}\right) \right)^2.
\end{align} 
where $n$ is the winding number. For simplicity, we focus on the case of $n=1$. Then, the envelope function $f(\varphi)$ has two zeros at
\be 
\varphi=\pi,\quad \varphi=\tan ^{-1}\left(\frac{1-b^2}{b^2+1},-\frac{2 b}{b^2+1}\right).
\ee 
By parametrizing $b$ as
\begin{equation}
b = \frac{1}{\tan\left(\frac{\varphi_0}{2}\right)},
\end{equation}
the second zero point becomes $\pi+\varphi_0$.  In the region $\varphi\in (\pi,\pi+\varphi_0)$, the vector is negative, while in the other regions, the vector field is positive. The zero point $\varphi=\pi $ acts as a sink, and the other zero point acts as a source. 
 Solving the stabilizer equation, we get
\begin{align}
u(\varphi) &= 4b^2 + \frac{2 + 8b^2}{F} - \frac{3}{F^2} + \frac{\fd - 1}{4} \frac{F^2}{(b + \tan\frac{\varphi}{2})^2 \cos^4\frac{\varphi}{2}},
\end{align} 
where $\fd$ is the orbit constant. To have some visualization of this geometry, we consider the case of $\fd=0$ and employ the Roberts mapping to transform the geometry into the \Poincare coordinates. First, we need define the new spatial variable $\tilde{\varphi}$ via \eqref{tphi}. Because $f$ has zeros, the definition of $\tilde{\varphi}$ is piecewise:
\be 
d\tilde{\varphi}=\frac{d\varphi}{f}\ \rightarrow\ \tilde{\varphi}=\begin{cases}
	 2b\varphi+\log(\tan\frac{\varphi}{2}-\tan\frac{\varphi_0+\pi}{2}), & \varphi\in[0,\pi) ; \\
	 2b\varphi+\log(-\tan\frac{\varphi}{2}+\tan\frac{\varphi_0+\pi}{2}), & \varphi\in(\pi,\pi+\varphi_0)\  \label{ptphi} ;\\
	 2b(\varphi-2\pi)+\log(\tan\frac{\varphi}{2}-\tan\frac{\varphi_0+\pi}{2})& \varphi\in (\pi+\varphi_0,2\pi]\ ,
\end{cases}
\ee  
where we properly shifted $\varphi$ to make sure $\tilde{\varphi}|_{\varphi=0}=\tilde{\varphi}|_{2\pi}$.
figure \eqref{tphi} explicitly illustrates how the transformation $\tilde{\varphi}=2b\varphi+\log|\tan\frac{\varphi}{2}-\tan\frac{\varphi_0+\pi}{2}|$  maps the spatical circle $S^1$ into two copies of $\mathbb{R}$. As a result, the corresponding Roberts mapping maps the original geometry into two (subregions of) \Poincare geometries denoted by $\mathcal{M}_p^\pm$, which are joined at the spatial infinities $(\text{the zeros of $f$}: \varphi=\pi,\pi+\varphi_0)$. Furthermore, within the domain $\varphi\in(\pi,\pi+\varphi_0)$, the mapping becomes monotonically decreasing, implying that the corresponding \Poincare geometry $\mathcal{M}_p^-$ has the opposite orientation. 
\begin{figure}[h]
\begin{center}
	 \includegraphics[scale=1]{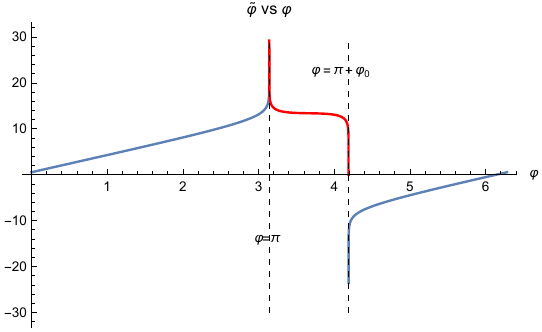}
  \caption{The relationship between 
$\tilde{\varphi}$ and $\varphi$ as defined in \eqref{ptphi} with $\varphi_0=\frac{\pi}{3}$.}\label{tphi}
\end{center}
\end{figure}

The effective length is now non-zero:
\begin{equation}
L_{\text{eff}} = \int_0^{2\pi} \frac{d\varphi}{f} = \int_0^{2\pi} \left( 2b + \frac{1}{b + b\cos\varphi + \sin\varphi} \right) d\varphi = 4\pi b \equiv \Delta. \label{i2}
\end{equation}
The holographic mass is also non-trivial:
\begin{equation}
M_{\text{holo}} = \frac{c}{12} \int_0^{2\pi} \frac{d\varphi}{2\pi} \left( \frac{\fd}{f} - \frac{f'^2}{f} \right) = \frac{c}{12} b (2\fd + 2 + b^2). \label{tnholomass}
\end{equation}
Details of this integral can be found in Appendix \ref{appendix:integral}. From the effective length, we define $f_0 = \frac{2\pi}{L_{\text{eff}}}$, then the vacuum state corresponds to $u_0 = -1$, implying
\begin{equation}
\fd = u_0 f_0^2 = -\left( \frac{2\pi}{4\pi b} \right)^2 = -\frac{1}{4b^2}.
\end{equation}

\begin{figure}[h]
\centering
\includegraphics[scale=0.4]{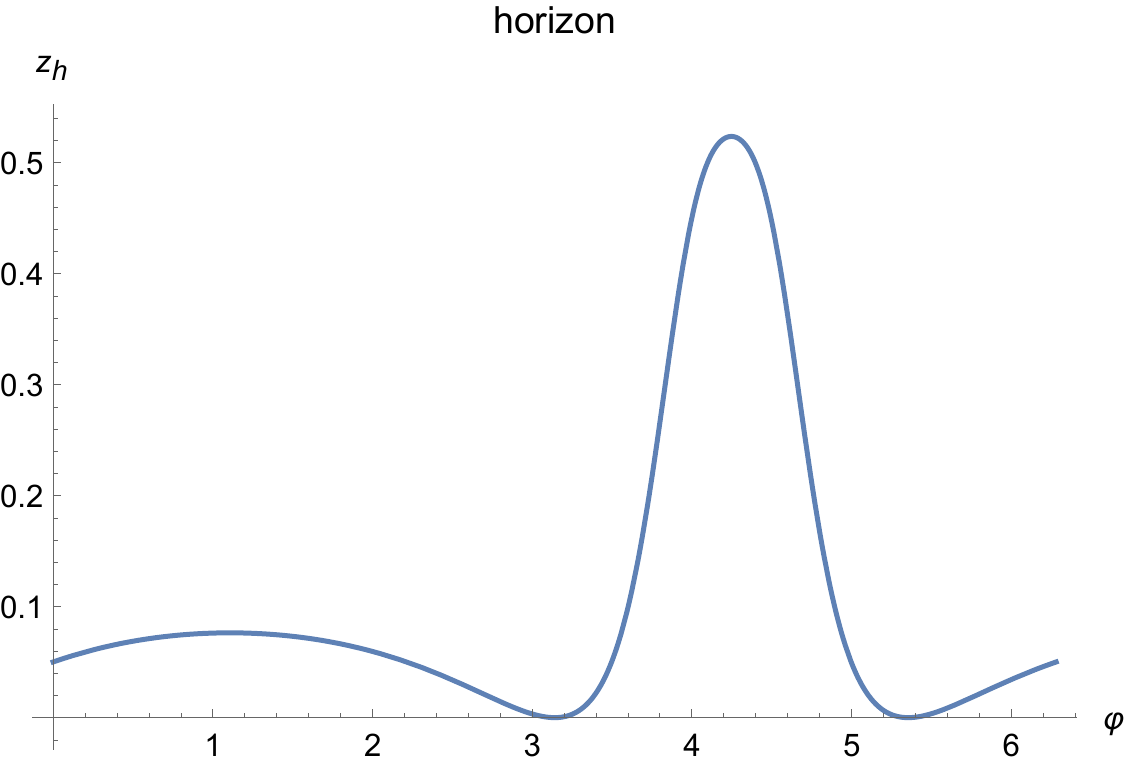}
\caption{The position of the horizon for $b = \frac{1}{2}, \fd = 20$.} \label{h2p}
\end{figure}

On the other hand, we find that when $\fd$ is sufficiently large, the smoothness conditions can be satisfied even though we do not find the explicit expression, placing the geometry in the black hole phase. An example of the black hole horizon is shown in figure \ref{h2p}.
The temperature and the entropy of the black hole are given by \eqref{tem} and \eqref{en}.

\section{KdV boundary condition}
\label{sec:kdv}
In this section, we extend our analysis to the KdV boundary conditions, a non-trivial extension of Dirichlet conditions, motivated by the integrable structure inherent to 2D CFTs \cite{Bazhanov:1994ft,Bazhanov:1996dr,Bazhanov:1998dq}. Beyond the infinite-dimensional Virasoro symmetries, 2D CFTs possess an infinite hierarchy of conserved KdV charges constructed from the energy-momentum tensor. The first few charges are:
\be 
Q_1\sim \int_0^{2\pi}d\varphi \ T,\quad Q_3\sim \int_0^{2\pi}d\varphi \ (TT),\quad Q_5\sim \int_0^{2\pi}d\varphi \left[ (T(TT))+\frac{c}{12}(\partial T)^2\right],
\ee 
where we have omitted the overall factor. Here, we will focus on the level-one KdV boundary condition and introduce the inhomogeneous Hamiltonian:
\be 
\mathcal{H}=\frac{1}{2\pi}(\mu u+u^2),\quad H=\int_0^{2\pi}\mathcal{H}d\varphi,
\ee 
where the parameter $\mu$ functions as a chemical potential. Crucially, $H$ cannot be expressed as a linear combination of Virasoro generators but instead as:
\begin{equation}
H_{\text{KdV}} = Q_3 + \mu Q_1. \label{eq:kdvh}
\end{equation}
Some properties, such as the conserved charges and black hole thermodynamics under KdV boundary conditions, have been investigated in \cite{Perez:2016vqo,Erices:2019onl,Dymarsky:2020tjh,Chen:2024ysb}. The bulk solution is expected to be dual to a quantum state (or thermal ensemble) characterized by non-trivial KdV charges. By ``non-trivial," we mean that the higher-level KdV charges are independent of $Q_1$. The quantum state with non-trivial KdV charges exhibits exotic thermalization behavior. It was argued in \cite{Cardy:2015xaa} and verified in \cite{Chen:2024lji,Chen:2024ysb} that local physical observables at late times are captured by the KdV Generalized Gibbs Ensemble (GGE). The holographic approach provides a straightforward setup for various computations. In this work, we will compute the holographic entanglement entropy of a subsystem directly using the RT formula, a task that has not been previously accomplished. In the context of inhomogeneous CFTs, the Hamiltonian \eqref{eq:kdvh} and its higher-level generalizations provide another class of solvable models that extend beyond all previously studied examples.

To obtain the bulk solution, we substitute the \eqref{def} to the stabilizer equation \eqref{stabEqn}. This leads to that $u(\varphi)$ needs to satisfy the (level-one) KdV equation
\be 
6uu'-4u'''=-\mu u'. \label{kdv}
\ee  
For general KdV Hamiltonians, this extends to generalized KdV equations. The complete family of solutions to these equations, known as finite-zone solutions, can be systematically constructed using integrability techniques \cite{Novikov:1974,Novikov:1984id,Belokolos:1994}. Some details about the solutions, termed as the one-zone solutions, to the level-one Hamiltonian \eqref{kdv} can be found in, for example, \cite{Dymarsky:2020tjh, Chen:2024ysb}. Here, we will not explain the construction but simply present the final result. Up to a constant $\varphi_c$, the solution is uniquely determined by three real-valued zone parameters $\{\lambda_i\}_{i=1}^3$, with the explicit form:
\begin{align}
u &= 4\left[-\partial_\varphi^2 \log\theta\big(U(\varphi-\varphi_c)|B\big) - c_u\right], \label{eq:solution} \\
U &= -\mathrm{i}\pi \frac{\sqrt{\lambda_3-\lambda_1}}{K(p)}, \quad p = \frac{\lambda_3-\lambda_2}{\lambda_3-\lambda_1}, \quad B = -2\pi \frac{K(1-p)}{K(p)},  \\
c_u &= -\lambda_3 + (\lambda_2-\lambda_1)\left(\frac{2\Pi(p,p)}{K(p)} - 1\right). \label{eq:cu-def}
\end{align}
Here $\theta[z|B] := \sum_{n\in\mathbb{Z}} e^{\frac{1}{2}Bn^2 + nz}$ denotes the Riemann theta function, $K(m)$ is the complete elliptic integral of the first kind, and $\Pi(m,n)$ the complete elliptic integral of the third kind\footnote{Mathematica implementations: \texttt{EllipticK[m]} and \texttt{EllipticPi[n,m]}.}. The chemical potential $\mu$ is also determined by the zone parameters as
\be 
\mu=-8(\lambda_1+\lambda_2+\lambda_3).
\ee 
The periodic boundary condition introduces quantization through the constraint:
\begin{equation}
U = \mathrm{i}k, \quad k \in \mathbb{Z}. \label{eq:quantization}
\end{equation}
Now, we derive some relevant properties of this solution. By integrating the equation of motion, we obtain:
\bea 
&&2u'^2=(u-4u_1)(u-4u_2)(u-4u_3),\quad u_1<u_2<u_3, \label{ude}\\ && u_1=\lambda_1+\lambda_2-\lambda_3,\quad u_2=\lambda_1+\lambda_3-\lambda_2,\quad u_3=\lambda_2+\lambda_3-\lambda_1.
\eea 
For the Hamiltonian given in \eqref{eq:kdvh}, the functional equation \eqref{def} simplifies to:
\begin{equation}
f = 2u + \mu. \label{kdvf}
\end{equation}
The differential equation \eqref{ude}, along with the relation \eqref{kdvf}, yields the following properties:
\begin{equation}
4u_1 \leq u \leq 4u_2, \quad 
-16\lambda_3 \leq f \leq -16\lambda_2. \label{f_range}
\end{equation}
Consequently, there are three possible scenarios:
\begin{enumerate}
	\item $f$ has no zeros, when $\lambda_3<0,f>0$ or $\lambda_2>0,f<0$. 
	\item $f$ has single zeros, when $\lambda_2<0<\lambda_3$.
	\item $f$ has double zeros, when $\lambda_2=0$ or $\lambda_3=0$.
\end{enumerate}
By adjusting $\lambda_3$, we can choose the solution to belong to different orbits. We can always apply a constant shift to eliminate the parameter $\varphi_c$ in the solution, meaning that the non-trivial parameters only correspond to the zone parameters of the one-zone solution. It is noteworthy that the orbit invariant is given by:
\begin{equation}
\fd = 2^{10} \lambda_1 \lambda_2 \lambda_3,
\end{equation}
which can be verified directly by substituting the solution \eqref{eq:solution} into \eqref{orbitconst}. 
Thus, we conclude that after fixing the chemical potential $\mu$ and the orbit constant $\fd $, there remains one degree of freedom. In fact, imposing periodic boundary conditions will quantize this one-dimensional space. The discrete parameter is denoted as $k$ in \eqref{eq:quantization}.  It is important to note that different solutions correspond to different envelope functions $f$, leading to distinct 2D boundary curved spacetimes. This is in contrast to the previous cases with Dirichlet boundary conditions.

To ensure the correct geometric orientation, we restrict our analysis to cases where $f>0$. The treatment of scenarios where $f$ contains zeros follows analogous arguments to those presented in the preceding section. Under the assumption that $f$ is positive-definite, we determine $f_0$ through:
\begin{align}
\frac{1}{f_0} &= \frac{1}{2\pi} \int_0^{2\pi} \frac{d\varphi}{f}
= \frac{k}{2\pi} \int_{4u_1}^{4u_2} \frac{\sqrt{2}du}{(u + \mu/2)\sqrt{(u - 4u_1)(u - 4u_2)(u - 4u_3)}}, \nonumber \\
&= -\frac{1}{2^4 \lambda_3} \frac{\Pi\left(\frac{\lambda_3 - \lambda_2}{\lambda_3}, p\right)}{K(p)}, \label{delta}
\end{align}
yielding the explicit expression:
\begin{equation}
f_0 = -\frac{2^4 \lambda_3 K(p)}{\Pi\left(\frac{\lambda_3 - \lambda_2}{\lambda_3}, p\right)}. \label{f0}
\end{equation}
Combining this result with $\fd=u_0f_0^2$, we derive the parameter $u_0$:
\begin{equation}
u_0 = \frac{\lambda_1 \lambda_2}{\lambda_3} \left( \frac{2 \Pi\left(\frac{\lambda_3 - \lambda_2}{\lambda_3}, p\right)}{K(p)} \right)^2. \label{u0}
\end{equation}
Crucially, these parameters must also satisfy the boundary condition:
\begin{equation}
f_0 - 2u_0 = \mu, \label{boundary}
\end{equation}
which imposes constraints on allowable orbits. In other words, this relationship implies that specific values of $u_0$ cannot be freely chosen--they require precise tuning of the chemical potential $\mu$ to maintain the consistency.

\begin{figure}[h]
\begin{center}
    \includegraphics[scale=1]{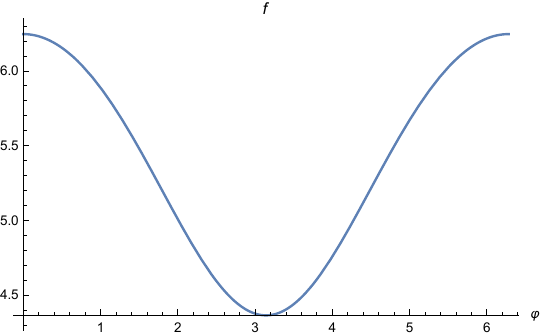}
    \caption{The envelope function $f$ for the solution with the parameters: $k = 1, \ \lambda_3 = -0.2728, \ p = 0.3744.$.} \label{f1}
\end{center}
\end{figure}

As demonstrated in \cite{Dymarsky:2020tjh}, the smoothness conditions cannot be satisfied when $f>0$, thereby precluding the existence of black-hole phases. We therefore focus on non-black-hole solutions characterized by negative values of $u_0$, which admit two distinct parameter configurations: 
\begin{align}
\lambda_1 &< 0 < \lambda_2 < \lambda_3, \label{c1} \\
\lambda_1 &< \lambda_2 < \lambda_3 < 0. \label{c2}
\end{align}
Imposing the $f>0$ requirement through \eqref{f_range}, we find that only configuration \eqref{c2} remains admissible. Similarly, we expect that the vacuum state solution corresponds to the choice $u_0=-1$. We need to identify the chemical potentials that permit this choice. Our approach proceeds as follows: First, we parameterize $\lambda_i$ in terms of $k$, $p$, and $\lambda_3$. Then, we employ the relationships between $\lambda_i$, $f_0$, and $u_0 = -1$ to derive the following two conditions:
\begin{align}
  2u_0 - f_0 &= 8 \left(-\frac{k^2 K(p)^2}{\pi^2} - \frac{k^2 p K(p)^2}{\pi^2} + 3 \lambda_3 \right), \\
  u_0 f_0^2 &= 1024 \lambda_3 \left(\lambda_3 - \frac{k^2 K(p)^2}{\pi^2}\right) \left(\lambda_3 - \frac{k^2 p K(p)^2}{\pi^2}\right).
\end{align}
Next, we determine the allowed values of $f_0$ and $k$ based on these two conditions, subject to the constraints $\lambda_3 < 0$ and $0 < p < 1$. For example, the two conditions can be visualized in the $(\lambda_3 - p)$ plane. If the two curves intersect, a solution exists. An illustration is shown in Figure \eqref{global}:
\begin{figure}[h]
\begin{center}
    \includegraphics[scale=0.8]{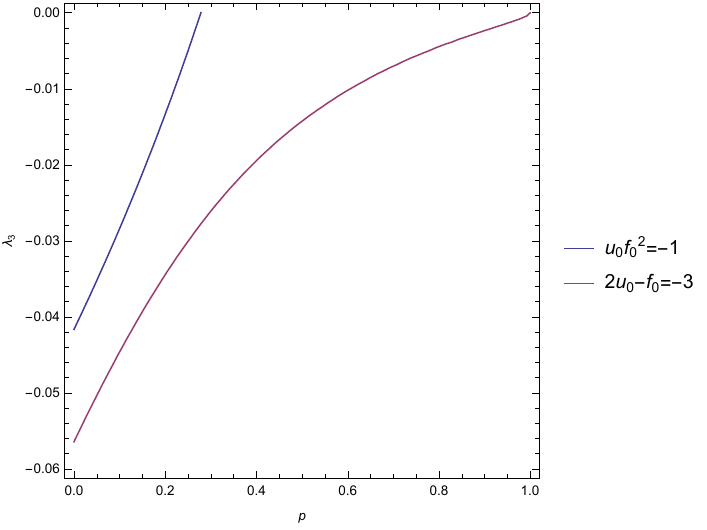}
    \caption{For $f_0 = 1$, we observe that no non-trivial solutions exist. In this example, $k = 1$.} \label{global}
\end{center}
\end{figure}
Using this method, we find that non-trivial solutions ($p \neq 0$) can only exist when the chemical potential satisfies:
\begin{equation}
  \mu \geq \frac{19}{2}.
\end{equation}
For instance, when $\mu = 10$, there exists only one solution:
\be 
k = 1, \quad \lambda_3 = -0.2728, \quad p = 0.3744.
\ee 
The corresponding envelope function $f$, visualized in figure \eqref{f1}, resembles a cosine-like function. When the chemical potential $\mu$ is close to the critical value $19/2$, the solution has a very small $p$ parameter. In this regime, we perform a perturbative expansion in the small-$p$ limit::
\be 
f=2\left(k^2 p \cos (k \varphi )+k^2 p-8 \lambda \right)|_{k=1}+\mathcal{O}(p^2),
\ee  
which remarkably coincides with $f_{\text{\Mobius}}$. This correspondence allows us to identify this solution as the holographic dual of the \Mobius deformation under KdV boundary conditions. Notably, the generalization to q-\Mobius deformations naturally emerges through solutions with wave number $k=q$ in the same perturbative regime.
As a concrete illustration, at $\mu = 20$ we find two distinct solutions:
\bea
&&k=1,\quad \lambda_3=-0.250098,\quad p=0.956111,\\
&&k=2,\quad \lambda_3=-0.306153,\quad p=0.314906.
\eea 

As discussed in Section \ref{sec:boundary}, under the KdV boundary condition, the Hamiltonian differs from the holographic mass. For the one-zone solutions, the Hamiltonian is expressed as in \eqref{eq:kdvh}, while the holographic mass defined in equation \eqref{eq:adm} is given by
\begin{equation}
M_{\text{holo}}[\varphi] = 2Q_3 + \mu Q_1 = H_{\text{KdV}} + Q_3,
\end{equation}
where $Q_1$ and $Q_3$ are also determined by the zone parameters:
\begin{align}
Q_1 &= \frac{c}{24\pi} \int_0^{2\pi} u \, d\varphi = \frac{c}{12} (-4c_u), \\
Q_3 &= \frac{c}{24\pi} \int_0^{2\pi} \left( u^2 - \frac{4}{3} u'' \right) d\varphi = \frac{\frac{c}{12} J - \mu Q_1}{3},
\end{align}
where $c_u$ is defined in \eqref{eq:cu-def}, and $J $ is defined as
\begin{equation}
J = 16 \left( \lambda_1^2 + (\lambda_2 - \lambda_3)^2 - 2\lambda_1(\lambda_2 + \lambda_3) \right).
\end{equation}
Assuming the RT formula is valid, the entanglement entropy of a subsystem $A\in(\varphi_1,\varphi_2)$ in the vacuum state is still given by this expression
\bea  
S_A&=&\frac{c}{6}\Gamma_A=\frac{c}{3}\log\frac{\frac{2}{f_0}\sin(\frac{f_0(\tilde{\varphi}_2-\tilde{\varphi}_2)}{2})}{\sqrt{(f(\varphi_1)f(\varphi_2))^{-1}}\epsilon},\\
\tilde{\varphi}_i&=&\int^{\varphi_i}\frac{d\varphi}{f}.\label{intphi}
\eea 
The integral \eqref{intphi} can be evaluated explicitly to obtain a fully analytic expression for the entanglement entropy. Due to the oscillatory nature of the function $f$, the mapping between $\tilde{\varphi}$ and $\varphi$ is not one-to-one. Considering the domain $\varphi \in [0,2\pi]$, we partition it into $2k$ subdomains:$\left[\frac{n\pi}{2k}, \frac{(n+1)\pi}{2k}\right], \, n = 0, 1, \dots, 2k-1$. Within each subdomain, the monotonicity of $f$ guarantees a one-to-one correspondence between $\tilde{\varphi}$ and $\varphi$. In particular, for even-indexed subdomains, $\tilde{\varphi}(\varphi)$ increases monotonically, while for odd-indexed subdomains, $\tilde{\varphi}(\varphi)$ decreases monotonically.
The effective length of each subdomain is given by 
\be 
\w=\frac{1}{2k}\frac{2\pi}{f_0}=\frac{\pi}{kf_0}.
\ee 
For $\varphi$ in the $2m$-th subdomain, we have
\begin{equation}
\tilde{\varphi} = 2m\omega + \int_{4u_1}^{u(\varphi)} \frac{\sqrt{2}\, du}{(2u + \mu)\sqrt{(u - 4u_1)(u - 4u_2)(u - 4u_3)}}, \label{even_sub}
\end{equation}
and for the $(2m-1)$)-th subdomain:
\begin{equation}
\tilde{\varphi} = 2m\omega - \int_{4u_1}^{u(\varphi)} \frac{\sqrt{2}\, du}{(2u + \mu)\sqrt{(u - 4u_1)(u - 4u_2)(u - 4u_3)}}. \label{odd_sub}
\end{equation}
The integral admits a closed-form expression:
\begin{equation}
\int_{4u_1}^{u(\varphi)} \frac{\sqrt{2}\, du}{(2u + \mu)\sqrt{(u - 4u_1)(u - 4u_2)(u - 4u_3)}} = -\frac{\Pi\left(1 - \frac{\lambda_2}{\lambda_3}; \gamma \big| p\right)}{128\lambda_1\lambda_2\sqrt{\lambda_3 - \lambda_1}}, \label{ellint}
\end{equation}
where $\Pi(n; \phi | m)$ denotes the incomplete elliptic integral of the third kind, and $\gamma $ is defined as
\begin{equation}
\gamma = \arcsin\left(\sqrt{\frac{u - 4u_1}{4u_2 - 4u_1}}\right) = \arcsin\left(\sqrt{\frac{u - 4(\lambda_1 + \lambda_2 - \lambda_3)}{8(\lambda_3 - \lambda_2)}}\right). \label{gamma_def}
\end{equation}

\section{Discussions}
\label{sec:conclusion}
We conclude by discussing several open questions and potential extensions of this work:

\textbf{Time-Dependent Spacetimes, Quantum Quenches, and Floquet CFTs:} While this work focused on static solutions, these are special cases of the general solutions \eqref{general} with constant $\mathcal{L}$ and $\bar{\mathcal{L}}$. Extending our analysis to time-dependent geometries would allow for the study of holographic quantum quenches in inhomogeneously deformed CFTs. For instance, previous works have explored local quantum quench \cite{Mao:2024cnm} in the presence of SSD deformations and argued equivalences between local joining quenches and \Mobius quenches \cite{Kudler-Flam:2023ahk}. Furthermore, since Floquet and driven CFTs can be understood as quenched systems, our approach could be applied to study their holographic duals. We will report this study in a future work.

\textbf{Boundary Effects and Inhomogeneously Deformed BCFTs:} This work and most previous studies have focused on 2D CFTs with periodic boundary conditions. However, as pointed out in \cite{Liu:2023tiq}, open boundary conditions introduce non-trivial boundary effects. Relaxing the periodicity constraint would enable the exploration of more general deformation functions. Studying inhomogeneously deformed boundary CFTs (BCFTs) in our holographic setup is a promising direction. Some works along this direction can be found in \cite{Das:2022pez,Bernamonti:2024fgx}

\textbf{3D C-Metric:} We revealed a connection between the Type-I 3D C-metric and the q-\Mobius deformation. This connection deserves further investigation. Additionally, we expect that Type-II and Type-III 3D C-metrics could also be realized within our holographic framework using non-periodic deformation functions. Another intriguing problem is constructing a 3D C-metric with non-trivial angular momentum, which might correspond to a q-\Mobius deformation where the chiral and anti-chiral sectors have different orbit constants. Insights from the Weyl-Fefferman-Graham gauge, as suggested in \cite{Arenas-Henriquez:2024ypo} for studying 3D metrics, could also be valuable for studying these geometries.

\textbf{$T_{n,\Delta}$ and $\tilde{T}_{n,\pm}$ Deformations in Lattice Models:} The SSD deformation was originally introduced in lattice models as a smooth boundary condition. Since we showed that the $\tilde{T}_{n,\pm}$ deformation can be understood as a perturbation of the SSD deformation, it is natural to expect that $\tilde{T}_{n,\pm}$ can also serve as a smooth boundary condition in lattice models. The $T_{n,\Delta}$ deformation, on the other hand, is more intriguing because it appears to glue multiple systems together in field theory. Investigating these deformations in lattice models is an exciting avenue for future research.

\textbf{Field Theoretic Analysis of KdV Hamiltonians:} We demonstrated that KdV Hamiltonians constitute a new solvable family of inhomogeneous Hamiltonians, with deformations that cannot be undone by simple conformal transformations. It would be interesting to study KdV Hamiltonians directly from the field theory perspective, potentially uncovering new insights into their physical significance and applications.

\section*{Acknowledgments}
We are grateful to Chen Bai, Weibo Mao, Farzad Omidi, Cheng Peng, Huajia Wang, Jie-qiang Wu, and Yu-Xuan Zhang for their insightful discussions and valuable perspectives. JT extends special appreciation to Mattias Gaberdiel for his critical questions and constructive feedback during the \textit{2024 KITS Retreat}. This work was supported through visiting scholar funding from the Kavli Institute for Theoretical Sciences (KITS) at the University of Chinese Academy of Sciences (UCAS).

\appendix
\section{Classification of the Virasoro Coadjoint Orbits}
\label{appendix:virasoro}

In this appendix, we review the classification of the coadjoint orbits of the Virasoro group, following \cite{Balog:1997zz} and \cite{Oblak:2016eij}.

\subsection{Diffeomorphisms of the Circle}
Let us consider the unit circle $S^1 = \{e^{\im \varphi} \in \mathbb{C} \,|\, \varphi \in [0, 2\pi)\}$. The circle can also be represented as the quotient $\mathbb{R}/(2\pi \mathbb{Z})$, where $\mathbb{R}$ serves as the universal cover of $S^1$, and $\varphi$ is a coordinate on $S^1$. The diffeomorphism group of $S^1$ factorizes as $\text{Diff}(S^1) = \text{Diff}^+(S^1) \oplus \text{Diff}^-(S^1)$, where $\text{Diff}^+(S^1)$ is the set of all orientation-preserving diffeomorphisms of the circle, defined by:
\be 
\text{Diff}^+(S^1) = \{\tilde{\varphi} \,|\, \partial_\varphi \tilde{\varphi} > 0, \quad \tilde{\varphi}(\varphi + 2\pi) = \tilde{\varphi}(\varphi) + 2\pi\}.
\ee 
Here, $\tilde{\varphi}$ can be interpreted as a new coordinate on $S^1$, and the map $\varphi \to \tilde{\varphi}$ represents the diffeomorphism from the old coordinate to the new one. The Lie algebra of $\text{Diff}(S^1)$ is spanned by infinitesimal diffeomorphisms:
\be 
\text{diff}(S^1) = \{f(\varphi) \,|\, \varphi \to \tilde{\varphi} = \varphi + \epsilon f(\varphi)\},
\ee 
which can be identified as the space of smooth vector fields on $S^1$, denoted as $\text{Vect}(S^1) = \{f(\varphi)\partial_{\varphi}\}$. The dual vector space is given by $\{u(\varphi)d\varphi^2 \,|\, u(\varphi) \in C^\infty(S^1)\}$, with the inner product defined as
\be 
\langle u, f \rangle = \frac{1}{2\pi} \int_0^{2\pi} u(\varphi) f(\varphi) d\varphi,
\ee 
which is invariant under diffeomorphisms.

\subsection{The Virasoro Group}
The Virasoro group, denoted as $\widehat{\text{Diff}}(S^1)$, is the universal central extension of $\text{Diff}(S^1)$. Its Lie algebra is the direct sum $\text{Vect}(S^1) \oplus \mathbb{R}$. The Virasoro adjoint vectors are pairs $(f(\varphi)\partial_\varphi, \lambda)$, while the dual vectors, also known as coadjoint vectors, are pairs $(u(\varphi)d\varphi^2, c)$. The centrally extended inner product is given by
\be 
\langle (u, c), (f, \lambda) \rangle = \frac{1}{2\pi} \int_0^{2\pi} u(\varphi) f(\varphi) d\varphi + \frac{c}{24} \lambda. \label{inner_product}
\ee 
Under the diffeomorphism $\varphi \to \tilde{\varphi}$, the vector field and the dual vector transform as
\bea 
&&\tilde{u}(\tilde{\varphi}(\varphi)) = \frac{1}{\tilde{\varphi}'^2} \big(u(\varphi) + 2\{\tilde{\varphi}, \varphi\}\big),  \\ 
&&\tilde{f}(\tilde{\varphi}(\varphi)) = \tilde{\varphi}' f(\varphi),  \label{tf}
\eea 
where $\{\cdot, \cdot\}$ denotes the Schwarzian derivative.

\subsection{Virasoro Coadjoint Orbits}
Fixing a coadjoint vector $u$, its orbit under the group action is defined as $W_u = \{\tilde{u}(\tilde{\varphi})\}$. A natural symplectic structure can be introduced on $W_u$. If the orbit is quantized, the resulting quantum Hilbert space corresponds to a unitary representation of the Virasoro group. Thus, classifying the unitary representations of the Virasoro group reduces to classifying its coadjoint orbits.

The stabilizer subgroup, $G(u)$, consists of diffeomorphisms $\tilde{\varphi}$ that leave $u$ invariant:
\be 
u(\tilde{\varphi}) = \tilde{u}(\tilde{\varphi}) = \frac{1}{\tilde{\varphi}'^2} \big(u(\varphi) + 2\{\tilde{\varphi}, \varphi\}\big).
\ee 
While solving this equation explicitly is challenging, the Lie algebra of the stabilizer can be analyzed. It is spanned by vector fields $f$ satisfying the linearized equation:
\be 
\delta_f u = f u' + 2 f' u - 2 f''' \equiv \mathcal{D}f.
\ee 
This is a third-order linear differential equation, so the stabilizer is at most three-dimensional, depending on the specific form of $u$. Integration of this equation yields a diffeomorphism-invariant constant:
\be 
u f^2 - 2 f f'' + (f')^2 = \fd. \label{appen:fd}
\ee 
The zeros of $f$ play a crucial role in classifying stabilizers. When $f$ has no zeros, a diffeomorphism $\varphi \to \varphi_0$ defined by
\be 
\frac{d\varphi_0}{f_0} = \frac{d\varphi}{f} \label{appen:diff}
\ee 
can transform $f$ to a constant $f_0$. Consequently, $u$ is also transformed into a constant $u_0$, implying that the orbit has a constant representative $u_0 \in W_u$. Other points on the orbit can be obtained by applying diffeomorphisms.

The solution to the linearized stabilizer equation,
\be 
\partial \big((u_0 - \partial^2) f\big) = 0,
\ee 
is three-dimensional:
\be 
f = c_1 \frac{e^{\sqrt{u_0}\varphi}}{\sqrt{u_0}} + c_2 \frac{e^{-\sqrt{u_0}\varphi}}{\sqrt{u_0}} + c_3.
\ee 
However, imposing periodicity reduces the space to one-dimensional $f = f_0$, except in cases where
\be 
f = e^{\pm \im n \varphi}, \quad u_0 = -n^2, \quad n \in \mathbb{Z}.
\ee 
Thus, when $f$ has no zeros, the stabilizer group is either $S^1$ or $PSL^{(n)}(2, \mathbb{R})$, depending on the value of $u_0$. 
In the second case, with special linear combinations of the basis vectors, the vector field $f$ may have zeros. For example:
\be 
f_n = \cos n\varphi, \quad \tilde{f}_n = 1 - \cos n\varphi.
\ee 
In such cases, the diffeomorphism \eqref{appen:diff} degenerates. The corresponding orbit may no longer have a constant representative. By considering small perturbations $u = -n^2 + \delta u$, two additional stabilizer groups, $T_{n,\Delta}$ and $\tilde{T}_{n,\pm}$, can be constructed. The representatives of these two groups are:
\be  
f_{T_{n,\Delta \equiv 4\pi b}}(\varphi) = \frac{2}{nF} \cos\left(\frac{n\varphi}{2}\right) \left(\sin\left(\frac{n\varphi}{2}\right) + \frac{b}{n} \cos\left(\frac{n\varphi}{2}\right)\right),
\ee 
\be 
f_{\tilde{T}_{n,\pm}}(\varphi) = \frac{1}{H_{n,q}} \sin^2\left(\frac{n\varphi}{2}\right),
\ee 
where 
\bea  
F(\varphi) &=& \cos^2\left(\frac{n\varphi}{2}\right) + \left(\sin\left(\frac{n\varphi}{2}\right) + \frac{2b}{n} \cos\left(\frac{n\varphi}{2}\right)\right)^2, \\
H_{n,q} &=& 1 + \frac{q}{2\pi} \sin^2\left(\frac{n\varphi}{2}\right), \quad q = \pm 1.
\eea   

\subsection*{The Refined Classification}
The refined classification of the Virasoro coadjoint orbits is achieved by relating the orbit to Hill's equation:
\be 
\psi''(\varphi) = \frac{1}{4} u(\varphi) \psi(\varphi),
\ee 
which is invariant under diffeomorphisms, provided:
\be
\tilde{\psi}(\tilde{\varphi}) = \psi(\varphi) \left(\frac{d\tilde{\varphi}}{d\varphi}\right)^{1/2}.
\ee 
Hill's equation is a second-order differential equation, so its solution space is two-dimensional, characterized by the monodromy matrix $M \in SL(2, \mathbb{R})$, defined as:
\bea 
&&\psi_1 \psi_2' - \psi_2 \psi_1' = -1, \\
&&\begin{bmatrix}
	\psi_1(\varphi+2\pi)\\
	\psi_2(\varphi+2\pi)
\end{bmatrix} = M
\begin{bmatrix}
	\psi_1(\varphi)\\
	\psi_2(\varphi)
\end{bmatrix},
\eea 
where $\psi_1$ and $\psi_2$ are normalized solutions of Hill's equation. 

The correspondence between the Virasoro coadjoint orbits and Hill's equation is twofold:
1) Given $u(\varphi)$, one can solve for $\psi_1$ and $\psi_2$, up to a change of basis, which amounts to an $SL(2, \mathbb{R})$ conjugation of $M$.
2) Given two normalized solutions $\psi_1$ and $\psi_2$, one can reconstruct $u(\varphi)$ and $f(\varphi)$. Thus, a Virasoro coadjoint orbit corresponds to a conjugacy class of $SL(2, \mathbb{R})$. However, different orbits may map to the same conjugacy class. To remove this degeneracy, an additional discrete parameter, the winding number, is introduced. The winding number is defined through the ratio:
\be 
\eta(\varphi) = \frac{\psi_1(\varphi)}{\psi_2(\varphi)},
\ee 
which transforms under $\text{Diff}(S^1)$ as a function with the property $\tilde{u}(\tilde{\varphi}) = 0$, where $\tilde{\varphi} = \eta$. One can interpret $\eta(\varphi)$ as a path on the circle. The winding number is the number of laps around the circle during the interval $\varphi \to \varphi + 2\pi$. To summarize, there exists a well-defined injective map from $W_u$ to $([M], n)$, where $[M]$ is the conjugacy class and $n$ is the winding number. The conjugacy classes of $SL(2, \mathbb{R})$ are classified as elliptic ($\text{Tr}(M) < 2$), parabolic ($\text{Tr}(M) = 2$), and hyperbolic ($\text{Tr}(M) > 2$). The refined classification is given by $(n \in \mathbb{N}, b > 0)$:
\be 
\{\mathcal{C}(\omega), \mathcal{B}_0(b), \mathcal{P}_0^{+}\} = S^1; \quad \mathcal{E}_n = PSL^{(n)}(2, \mathbb{R}); \quad \mathcal{B}_n(b) = T_{n, \Delta = 4\pi b}; \quad \mathcal{P}_n^\pm = \tilde{T}_{n, \pm}.
\ee  
The class $\mathcal{C}(\omega)$ corresponds to elliptic monodromy parameterized by the matrix:
\be 
\begin{pmatrix}
	\cos(2\pi \omega) & \sin(2\pi \omega) \\
	-\sin(2\pi \omega) & \cos(2\pi \omega)
\end{pmatrix}, \quad \omega \in (0, \frac{1}{2}) \cup (\frac{1}{2}, 1).
\ee 
The stabilizer group is $G(\mathcal{C}(\omega)) = U(1) = S^1$, the rotation group. The winding number is obtained by $n = [2/f_0]$.

The classes $\mathcal{B}_0(b)$ and $\mathcal{B}_n(b)$ correspond to hyperbolic monodromy parameterized by the matrix:
\be 
\begin{pmatrix}
	e^{2\pi b} & 0 \\
	0 & e^{-2\pi b}
\end{pmatrix}, \quad b > 0,
\ee
and have winding numbers 0 and $n$, respectively. The stabilizer group is $G(\mathcal{B}_\eta(b)) = \mathbb{R}^\times$, the dilation group.

The classes $\mathcal{P}_0^{+}$ and $\mathcal{P}_n^\pm$ correspond to non-degenerate parabolic monodromy represented by the matrices:
\be 
\pm \begin{pmatrix}
	1 & 0 \\
	q & 1
\end{pmatrix}, \quad q = \pm 1,
\ee 
and have winding numbers 0 and $n$. The stabilizer group is the maximal nilpotent subgroup:
\be 
\begin{pmatrix}
	\pm 1 & 0 \\
	\gamma & \pm 1
\end{pmatrix}, \quad \forall \gamma \in \mathbb{R}. \label{nonpara}
\ee 

The class $\mathcal{E}_n$ corresponds to degenerate parabolic monodromy, represented by the matrices:
\be 
\pm \begin{pmatrix}
	1 & 0 \\
	0 & 1
\end{pmatrix},
\ee 
and has winding number $n$. The stabilizer group is $PSL^{(n)}(2, \mathbb{R})$, the $n$-\Mobius transformation group.

\section{General AdS$_3$ solutions}
\label{appendix:general}
The general asymptotic AdS$_3$ solution can be written in the FG gauge
\be 
ds^2=g_{\alpha\beta}(z,x^\alpha)dx^\alpha dx^\beta+\frac{dz^2}{z^2},\quad g_{\alpha\beta}(z,x^\alpha)=\frac{1}{z^2}g_{\alpha\beta}^{(0)}+g_{\alpha\beta}^{(2)}+z^2 g_{\alpha\beta}^{(4)}.
\ee 
If we fix the asymptotic boundary metric to be
\be 
g_{ij}^{(0)}dx^i dx^j=e^\phi dw d\bar{w},
\ee 
then the whole 3d metric is given by
\bea 
&&ds^2=\frac{dz^2}{z^2}+\frac{1}{z^2}e^\phi dwd\bar{w}+\mathcal{T}dw^2+\bar{\mathcal{T}}d\bar{w}^2+R_\phi dwd\bar{w},\nn
&&\qquad +z^2 e^{-\phi}(\mathcal{T}dw+\frac{1}{2}R_\phi d\bar{w})(\bar{\mathcal{T}}d\bar{w}+\frac{1}{2}R_\phi d{w}),\label{general}
\eea 
with
\bea 
&&\mathcal{T}=\frac{1}{2}\partial_w^2 \phi-\frac{1}{4}(\partial_w \phi)^2+\mathcal{L}(w),\quad \bar{\mathcal{T}}=\frac{1}{2}\partial_{\bar{w}}^2 \phi-\frac{1}{4}(\partial_{\bar{w}} \phi)^2+\bar{\mathcal{L}}(\bar{w}),\label{llbar}\\
&& R_\phi=\partial_w\partial_{\bar{w}}\phi.
\eea 
$\mathcal{L}(w)$ and $\bar{\mathcal{L}}(\bar{w})$ are two arbitrary functions that characterize different states of the holographic CFT. This bulk solution is a generalization of the \Banados geometry. Locally, this geometry can be transformed into the \Poincare AdS$_3$ solution
\be 
ds^2=\frac{d\eta^2+dYd\bar{Y}}{\eta^2},\label{poincare}
\ee 
via the mapping:
\bea 
&&\frac{1}{\eta}=\frac{1}{z}\(\frac{e^\phi}{(\partial_w y\partial_{\bar{w}}\bar{y} )}\)^{\frac{1}{2}}+\frac{z}{4}\(\frac{e^\phi}{(\partial_w y\partial_{\bar{w}}\bar{y} )}\)^{-\frac{1}{2}}\(\frac{(\partial_{\bar{w}}^2\bar{y}-\partial_{\bar{w}}\bar{y}\partial_{\bar{w}}\phi)(\partial_{{w}}^2{y}-\partial_{{w}}{y}\partial_{{w}}\phi)}{(\partial_w y\partial_{\bar{w}}\bar{y} )^2}\),\nn
&&Y=y(w)-\frac{2(\partial_w y)^2}{\partial^2_wy+\partial_w y\(\frac{4e^\phi \partial_{\bar{w}}\bar{y}}{z^2(\partial_{\bar{w}}^2\bar{y}-\partial_{\bar{w}}\bar{y}\partial_{\bar{w}}\phi)}-\partial_w \phi\)},\nn
&&\bar{Y}=\bar{y}(\bar{w})-\frac{2(\partial_{\bar{w}} \bar{y})^2}{\partial^2_{\bar{w}}\bar{y}+\partial_{\bar{w}} \bar{y}\(\frac{4e^\phi \partial_{{w}}{y}}{z^2(\partial_{{w}}^2{y}-\partial_{{w}}{y}\partial_{{w}}\phi)}-\partial_{\bar{w}} \phi\)}. \label{mapping}
\eea 
Correspondingly, the two functions are given by
\be 
\mathcal{L}=-\frac{1}{2}\{y(w),w\}=\frac{3y''^2-2y'y'''}{4y'^2},\quad \bar{\mathcal{L}}=-\frac{1}{2}\{\bar{y}(\bar{w}),\bar{w}\}=\frac{3\bar{y}''^2-2\bar{y}'\bar{y}'''}{4\bar{y}'^2}.
\ee

\section{Details of some integrals}
\label{appendix:integral}
In this appendix, we present the details of some complicated integrals. Let us first show \eqref{ssdleff}:
\be 
I_1=\int_0^{2\pi}\frac{\d\varphi}{ k\sin(\frac{\varphi}{2})\sin(\frac{(\varphi-\varphi_0)}{2})}=0,
\ee 
which can be easily verified by Mathematia. Here, we will show this in a more general method which can be used in other cases. The idea is to use the residue theorem. Introducing $\varphi=-\im\log u$, we can rewrite $I_1$ as a loop integral
\be 
I_1=-\oint \frac{4\im e^{\frac{\im \varphi_0}{2}}\d u}{k(u-e^{\im \varphi_0})(u-1)}=\pi\im \(\text{Res}(u,1)+\text{Res}(u,e^{\im \varphi_0})\)=0\, .
\ee 
Next, let us show \eqref{i2}:
\bea  
I_2 &=&\int_0^{2\pi}\d\varphi\(\frac{2 b^2 \cos (\varphi )+2 b^2+2 b \sin (\varphi )+1}{b \cos (\varphi )+b+\sin (\varphi )}\),\\
&=&\oint \(\frac{b-i}{(b-i) u+b+i}-\frac{2 i b}{u}-\frac{1}{u+1}\)\d u ,\\
&=&2\pi\im\(\frac{1}{2}\text{Res}(u,-1)+\frac{1}{2}\text{Res}(u,\frac{\im+b}{\im-b})+\text{Res}(u,0)\)=2\pi\im (-2\im b)=4\pi b. 
\eea  
Similarly, let us compute the integral \eqref{tnholomass}:
\be 
I_3=\int_0^{2\pi}\frac{\d\varphi}{2\pi}\(\frac{d}{f}-\frac{f'^2}{f^2}\)=2bd-I_4,
\ee 
where
\bea  
I_4&=&\int_0^{2\pi}\frac{\d\varphi}{2\pi}\frac{f'^2}{f^2}=\frac{2 \sec \left(\frac{\varphi}{2}\right) \sin ^5\left(\frac{{\varphi_0}}{2}\right) \sin ^2\left(\varphi-\frac{\varphi_0}{2}\right) \sec \left(\frac{\varphi-\varphi_0}{2}\right)}{\pi  (2 \cos (\varphi-{\varphi_0})+2 \cos (\varphi)+\cos ({\varphi_0})+3)^3},\\
&=&\oint\(\frac{u \left(-1+e^{i \varphi _0}\right){}^5 \left(-u^2+e^{i \varphi _0}\right){}^2}{2 \pi  (u+1) \left(u+e^{i \varphi _0}\right) \left(2 u+e^{i \varphi _0}+1\right){}^3 \left(u+(u+2) e^{i \varphi _0}\right){}^3}\)\d u,\\
&=&2\pi \im \, \text{Res}(u,-\frac{1+e^{\im \varphi_0}}{2})=\frac{1}{2} \left(\cos \left(\varphi _0\right)-3\right) \cot \left(\frac{\varphi _0}{2}\right) \csc ^2\left(\frac{\varphi _0}{2}\right),\\
&=&-b \left(b^2+2\right).
\eea  

\bibliographystyle{unsrturl}
\bibliography{ref}

\end{document}